\documentclass[sigconf,nonacm,10pt]{acmart}

\makeatletter
\def\@ACM@checkaffil{
    \if@ACM@instpresent\else
    \ClassWarningNoLine{\@classname}{No institution present for an affiliation}%
    \fi
    \if@ACM@citypresent\else
    \ClassWarningNoLine{\@classname}{No city present for an affiliation}%
    \fi
    \if@ACM@countrypresent\else
        \ClassWarningNoLine{\@classname}{No country present for an affiliation}%
    \fi
}
\makeatother

\usepackage{graphicx}
\usepackage{url}
\usepackage{amsmath}
\usepackage{bm}
\usepackage{xcolor}
\usepackage{colortbl}
\usepackage{soul}
\usepackage[linesnumbered,ruled,vlined]{algorithm2e}
\usepackage[export]{adjustbox}
\usepackage{hyperref}
\usepackage{rotating}
\usepackage{array}
\usepackage{tabularx}
\usepackage{tikz}
\usepackage{float}
\usepackage[noabbrev, capitalise]{cleveref}
\usepackage{multirow}
\usepackage{wrapfig}
\usepackage[font=footnotesize,labelfont=bf]{caption, subcaption}
\usepackage{caption}
\usepackage{listings}
\definecolor{brightmaroon}{rgb}{0.76, 0.13, 0.28}
\setul{1pt}{0.5pt} 

\begin{document}


\title {\LARGE Real-Time Mobile Video Analytics for Pre-arrival Emergency Medical Services}




\author{\large Liuyi Jin, Amran Haroon, Radu Stoleru, Pasan Gunawardena, Michael Middleton*, Jeeeun Kim}

\affiliation{\small Computer Science and Engineering, Texas A\&M University \\ *Emergency Medical Services (EMS), Texas A\&M University }

\email{{liuyi, amran.haroon, stoleru, pgunawardena, mmiddleton, jeeeun.kim}@tamu.edu}


%


\begin{abstract}

Timely and accurate pre-arrival video streaming and analytics are critical for emergency medical services (EMS) to deliver life-saving interventions. Yet, current-generation EMS infrastructure remains constrained by one-to-one video streaming and limited analytics capabilities, leaving dispatchers and EMTs to manually interpret overwhelming, often noisy or redundant information in high-stress environments. We present \textbf{TeleEMS}, a mobile live video analytics system that enables pre-arrival multimodal inference by fusing audio and video into a unified decision-making pipeline before EMTs arrive on scene.

TeleEMS comprises two key components: \textbf{TeleEMS Client} and \textbf{TeleEMS Server}. The TeleEMS Client runs across phones, smart glasses, and desktops to support bystanders, EMTs en route, and 911 dispatchers. The TeleEMS Server, deployed at the edge, integrates EMS-Stream, a communication backbone that enables smooth multi-party video streaming. On top of EMSStream, the server hosts three real-time analytics modules: (1) audio-to-symptom analytics via EMSLlama, a domain-specialized LLM for robust symptom extraction and normalization; (2) video-to-vital analytics using state-of-the-art rPPG methods for heart rate estimation; and (3) joint text–vital analytics via PreNet, a multimodal multitask model predicting EMS protocols, medication types, medication quantities, and procedures.

Evaluation shows that EMSLlama outperforms GPT-4o (exact-match 0.89 vs. 0.57) and that text–vital fusion improves inference robustness, enabling reliable pre-arrival intervention recommendations. TeleEMS demonstrates the potential of mobile live video analytics to transform EMS operations, bridging the gap between bystanders, dispatchers, and EMTs, and paving the way for next-generation intelligent EMS infrastructure.

\end{abstract}

\maketitle
\pagestyle{plain}

\balance

\section{Introduction}
\label{sec:intro}

\textbf{Pre-arrival emergency medical service (EMS)} refers to the EMS delivered in the critical time interval between a 911 call and the arrival of emergency medical technicians (EMTs). Outcomes in cardiac arrest, airway obstruction, and severe bleeding hinge on such pre-arrival time intervals~\cite{bystanderMildWait2007}. In the US, EMTs respond to over 37 million calls annually~\cite{nemsis2015,NEMSIS2023Report}. The average response times are approximately 7 minutes nationwide, rising to 13 minutes in rural areas, and nearly 10\% of rural cases face delays approaching 30 minutes. These extended medical response times have been directly associated with worse outcomes in trauma patients~\cite{responsetime2017,ResponseTimeTexas2021,increasedEMSTime2007}. Recognizing the life-saving potential of immediate intervention, governments have launched initiatives such as \textit{Until Help Arrives} and \textit{Call=Care}~\cite{UHA2025,CallCare2025,CallBoard2023}. These campaigns share a vision: a 911 call should not only dispatch EMTs but also activate EMT-guided, bystander-delivered care coordinated with dispatchers and en route teams. Realizing this vision requires infrastructures that favor more reliable, coordinated, and intelligent pre-arrival emergency care.

Current-generation EMS infrastructures, however, largely restrict communication to unidirectional, audio-based information flows from bystander to dispatcher to EMTs en route. Dispatchers must simultaneously coach bystanders, assess severity, decide on escalation, and relay updates forward. EMTs, in turn, prepare under time pressure based on the dispatcher’s interpretation accuracy and timeliness. This linear, audio-heavy pipeline imposes high cognitive loads and is prone to errors when information is noisy, incomplete, or rapidly evolving.


\begin{figure}[h]
    \includegraphics[width=\linewidth]{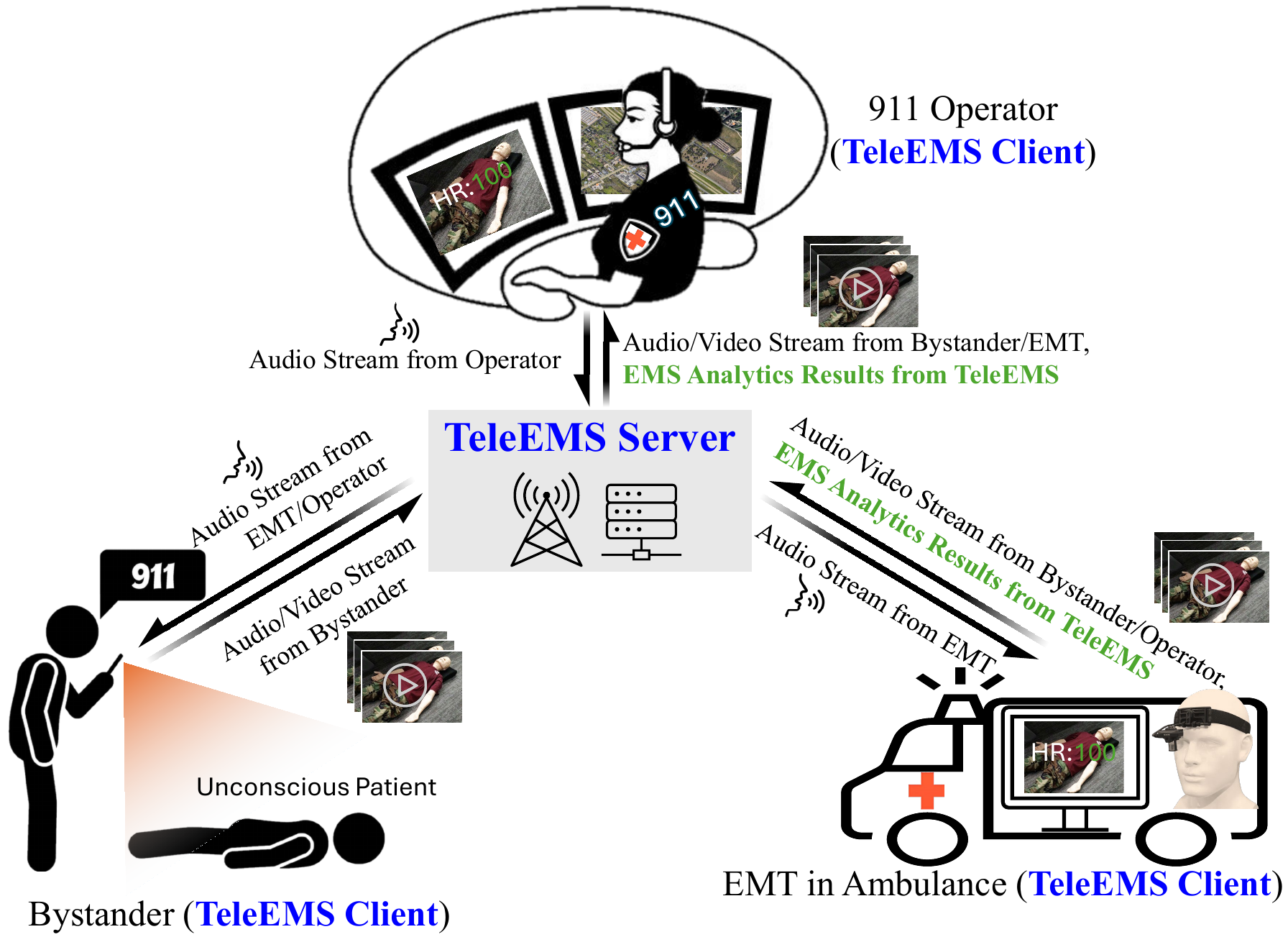}
\caption{TeleEMS application scenario example. The TeleEMS Server runs in the data center or on host machines of the cellular Core network, connecting three TeleEMS Clients: a bystander witnessing an unconscious patient, a 911 operator, and a dispatched EMT in the ambulance.}
    \Description{TeleEMS application scenario.}
    \label{fig:application_scenario}
\end{figure}

To relieve this burden and improve information consistency, \textit{next-generation pre-arrival EMS} must have three capabilities: (1) support multi-party participation with the true bidirectional audio and video streaming capability; (2) retain, structure, and display critical information from live streams with high fidelity; and (3) accurately recommend clinical intervention actions. No existing EMS platform provides this level of integrated, intelligent support. To address this gap, we propose \textbf{TeleEMS} (Figure~\ref{fig:application_scenario}), the first mobile live video analytics system designed to advance pre-arrival EMS and unlock the \textit{next-generation intelligent emergency medical response infrastructure}.

TeleEMS spans three client endpoints: the bystander’s smartphone, the dispatcher’s workstation, and the EMT’s smart glasses. These endpoints are linked through a dedicated edge server. Once a 911 video call begins, the bystander streams live video while speaking with both dispatcher and EMTs. The dispatcher, a certified EMT, evaluates urgency, locates the incident, decides on escalation, and shares real-time updates with EMTs in the ambulance. Inside the ambulance, EMTs monitor the scene through smart glasses and communicate seamlessly with both parties.

TeleEMS enables full-duplex audio across participants and streams video from the bystander to both dispatcher and EMT, ensuring that all stakeholders share a consistent audio-visual context throughout the pre-arrival window. Unlike current EMS infrastructures, TeleEMS introduces low-latency, reliable communication via an edge server co-located with the cellular core or local data centers. This edge platform not only supports robust connectivity in disaster-prone scenarios but also hosts intelligent video analytics to aid decision-making during the call. Together, the three capabilities discussed above advance pre-arrival EMS towards a more reliable, coordinated, and intelligent next-generation emergency medical response infrastructure.

The data, code, and model in TeleEMS will be open-sourced upon paper acceptance.

\section{Challenges and Contributions}
\label{sec:challenge_contribution}

TeleEMS advances pre-arrival EMS by addressing four critical gaps in current infrastructures and offering corresponding contributions:

\textbf{Challenge \#1}: \textit{Lack of fully integrated, multi-party, bidirectional video and audio streaming}.
Widely deployed EMS platforms (e.g., Carbyne’s c-Live Video~\cite{CARBYNE2025}, Prepared911~\cite{Prepared911}, RapidSOS~\cite{RapidSOS2025}, and Motorola Solutions CommandCentral Aware~\cite{MotorolaCommandCentralAware2025}) only stream the bystander’s video to dispatchers. EMTs en route receive filtered updates via dispatcher relays, losing crucial visual details and often introducing inaccurate information relays. Some newer systems, such as RapidDeploy’s Lightning~\cite{RapidDeploy2025}, attempt to forward videos to EMTs' smartphones directly. However, their audio information in the scene still depends on the dispatchers' accurate relay, preventing EMTs from guiding bystanders through urgent, scene-specific interventions like CPR or bleeding control. Moreover, these systems are proprietary, restricting academic participation and innovation.

\textbf{Contribution \#1}: \textit{EMS-Stream --- open-source, multi-party streaming}.
We design and implement EMS-Stream, the first open-source, cross-platform system supporting multi-party, bidirectional video and audio streaming among bystanders, dispatchers, and EMTs. During a 911 call, any participant can stream video or audio to all others, which is a critical capability absent in today’s EMS systems: unidirectional bystander video to both dispatcher and EMT, combined with bidirectional EMT–bystander audio. This allows EMTs to directly guide bystanders in delivering life-saving interventions before arrival. EMS-Stream supports dynamic joins/leaves, heterogeneous hardware devices (smartphones, dispatch consoles, smart glasses). This flexibility is particularly valuable when additional parties are involved beyond dispatchers and EMTs en route, such as remote medical specialists, language interpreters, or law enforcement officers coordinating scene safety. By open-sourcing EMS-Stream as a part of TeleEMS, we overcome functional gaps and provide the research community with an extensible testbed for advancing real-time EMS communication technology.

\textbf{Challenge \#2}: \textit{Absence of accurate, real-time extraction and normalization of medical symptoms}.
In current platforms (e.g., RapidSOS~\cite{RapidSOS2025}, Prepared911~\cite{Prepared911}), the key information extraction purely relies on raw speech-to-text transcriptions from noisy, high-stress calls. Recognition errors (e.g., ``arts'' for ``ards'') and lack of automated symptom extraction force dispatchers and EMTs to track symptoms manually. This adds substantial cognitive load during 911 calls, when attention is already stretched across triage, coordination, and real-time decision-making. Prior medical named entity recognition systems target clean, written text~\cite{scispacy2019medner}, leaving real-time symptom extraction and normalization from error-prone real-world EMS transcription texts unexplored.

\textbf{Contribution \#2}: \textit{EMSLlama --- real-time symptom extraction and normalization}.
We propose EMSLlama, a TeleEMS-integrated pipeline that fine-tunes Llama 3-8B~\cite{grattafiori2024llama3herdmodels,llama8bhugginface}, a state-of-the-art large language model (LLM), on a bystander–EMT conversation dataset created through a novel synthesis and augmentation technique, which reflects real-world EMS transcription noise. EMSLlama extracts symptom mentions from noisy speech-to-text transcriptions (i.e., transcriptions with incorrectly transcribed texts) and normalizes them into standardized terminology (e.g., ``arts'' $\rightarrow$ ``ards'') in real time. Normalized symptoms are displayed on dispatcher consoles and EMT smart glasses, persist across the call. This structured record reduces reliance on human memory, lowers cognitive burden, and improves continuity of care. Beyond symptoms, EMSLlama’s methodology can be generalized to other types of EMS key information extraction, such as the mechanism of injury.

\textbf{Challenge \#3}: \textit{Lack of integrated live video analytics}.
Although audio transcription is common, current platforms neglect automated analysis of live bystander video, which often contains vital patient and scene cues (e.g., cyanosis, chest rise, uncontrolled bleeding, burns, fire or smoke hazards). These must be manually interpreted, risking delays and missed interventions.


\textbf{Contribution \#3}: \textit{Extensible live video analytics in TeleEMS}.
We integrate real-time video analytics, beginning with remote photoplethysmography (rPPG) to estimate heart rate from patients' subtle facial pixel variations using TSCAN~\cite{tscan2020neurips,rppgtoolbox2023neurips}, a state-of-the-art rPPG model. Bystanders simply need to orient their phone camera toward the patient’s face. TeleEMS is modular, supporting future live video analytics integration such as bleeding detection, respiratory distress recognition, burn assessment, or hazard identification. This modular design enables TeleEMS to grow beyond rPPG toward a comprehensive real-time video intelligence system for pre-arrival EMS delivery.

\textbf{Challenge \#4}: \textit{Lack of actionable pre-arrival intervention recommendations}.
Dispatchers juggle communication, triage, and coordination; EMTs prepare en route. Deciding what bystanders should do before EMT arrival is an added cognitive burden requiring integration of incomplete multimodal data (audio, video, transcripts). Prior systems such as CognitiveEMS~\cite{CognitiveEMS2024IoTDI,Shu2019ABT,ShuICCPSWorkshop2018,Preum2019CognitiveEMSAC} and EMSAssist~\cite{EMSAssist2023MobisysJin,EMSAssistDemo2023} explored protocol recommendations. However, they focus on post-arrival contexts with complete audio information, ignore video analytics, and remain too general (e.g., protocols without dosage- or patient-specific detail).

\textbf{Contribution \#4}: \textit{PreNet — multimodal, patient-specific pre-arrival recommendations}.
We present PreNet, the first model to translate incomplete, pre-arrival multimodal data into actionable, patient-specific interventions. PreNet consumes textual inputs and vitals and outputs four recommendations: (1) the EMS protocol, (2) the specific medicine, (3) the medicine dosage, and (4) the procedures. These recommendations are displayed on dispatcher consoles and EMT smart glasses, offering concrete guidance that can be relayed to bystanders. This reduces decision burden, accelerates initiation of critical care.

\section{System design}
\label{sec:presceneems_design}

\begin{figure*}
    \includegraphics[width=\linewidth]{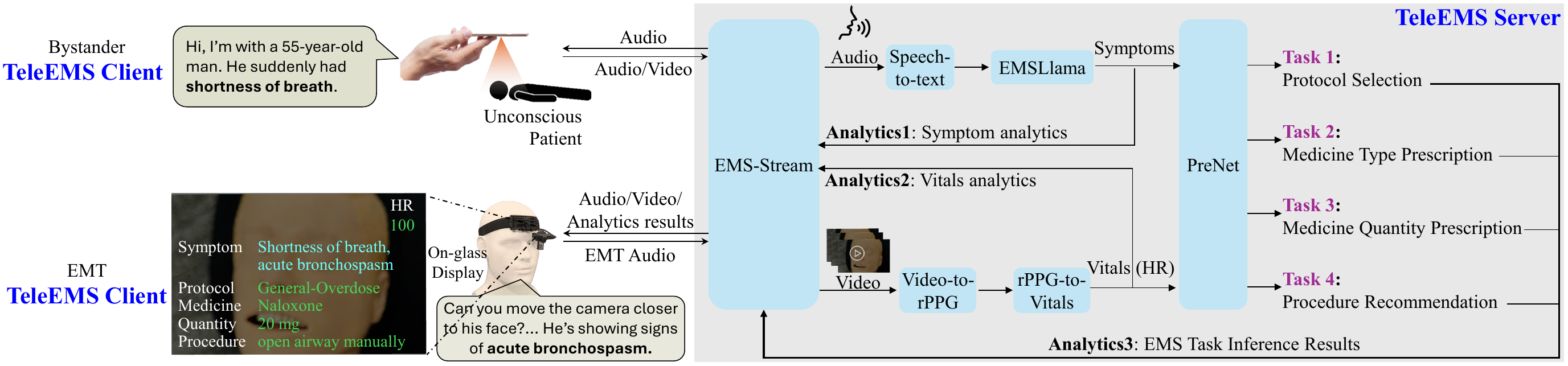}
\caption{Overview of TeleEMS system architecture.}
\Description{TeleEMS_system_design}
    \label{fig:system_design}
\end{figure*}

\subsection{TeleEMS system architecture}
As illustrated in Figure~\ref{fig:system_design}, TeleEMS is a mobile, live video analytics system designed to deliver real-time, actionable pre-arrival emergency medical support by integrating multi-party communication and multimodal multitask analytics into a unified architecture. The system comprises two main subsystems: \textbf{TeleEMS Clients}, deployed on bystander smartphones and EMT devices, and the \textbf{TeleEMS Server}, which performs analytics and decision support. Communication between clients and server is coordinated through \textbf{EMS-Stream}, a cross-platform streaming and data exchange module that supports simultaneous multi-party multimedia streaming, including text, audio, and video.
 
On the client side, bystanders initiate a 911 video call using the TeleEMS mobile application, which streams both audio and video from the incident scene to the server via EMS-Stream. EMTs en route, equipped with smart glasses running the TeleEMS client, receive live scene audio, video, and real-time analytics results overlaid on their on-glass displays. The bidirectional audio channel enables EMTs to directly instruct bystanders in performing life-saving interventions before arrival. While Figure~\ref{fig:system_design} depicts two clients (a bystander and an EMT en route), TeleEMS supports additional participants across heterogeneous hardware, such as dispatchers using desktop workstations (Figure~\ref{fig:application_scenario}). The TeleEMS Client supports dynamic participant management, enabling dispatchers, EMTs, and other responders to join or leave without interrupting the session.

The TeleEMS Server hosts three analytics pipelines. \textbf{Analytics 1} processes audio streams through a speech-to-text module followed by EMSLlama, which extracts and normalizes key medical symptoms from noisy transcriptions. \textbf{Analytics 2} processes video streams for vitals estimation: when instructed, bystanders orient the camera toward the patient’s face, enabling a video-to-vitals pipeline that estimates heart rate using the TSCAN model~\cite{rppgtoolbox2023neurips,tscan2020neurips}.\textbf{ Analytics 3} fuses normalized symptoms and vitals into our multimodal multitask model \textbf{PreNet} that performs four tasks: (1) EMS protocol selection, (2) medicine type prescription, (3) medicine quantity prescription, and (4) EMT procedure recommendation. Among the four, tasks 2-4 recommend specific actionable interventions. Outputs from these three analytics pipelines are aggregated by the EMS-Stream module and delivered in real time to EMT smart glasses, providing precise, actionable guidance that can be immediately relayed to bystanders.

By combining multi-party live streaming, multimodal analytics, and AI-driven decision support, TeleEMS transforms raw, unstructured audio and video into structured, actionable clinical information (symptoms and vitals) and intervention recommendations (tasks 1-4) during the pre-arrival phase. This integration reduces cognitive load for dispatchers and EMTs, improves situational awareness, and enables earlier initiation of life-saving interventions. With the components described above integrated, TeleEMS establishes a flexible and extensible foundation for next-generation emergency medical response infrastructure, enabling the integration of new analytics modules and communication capabilities over time.

\subsection{Edge server deployment rationale}

A core design choice in TeleEMS is to deploy the server on a private local edge node rather than in the cloud. This decision is grounded in the realities of large-scale emergencies, where public internet connectivity may be unreliable or unavailable while local cellular coverage remains intact~\cite{EMSAssist2023MobisysJin,EMSAssistDemo2023,EdgeCore,EdgeCoreWorkshop2022,DistressNetNG2024TCPS,AMVP2020SEC}. For example, during Hurricane Sandy, many New York City cell towers maintained local coverage, but their backhaul connections to the internet were severed by flooding, leaving devices with signal bars yet unable to access cloud services~\cite{BytagigNoInternet2025}. Similarly, during the Boston Marathon bombing, local networks were overwhelmed, and only emergency services retained prioritized uplink access; civilians could connect to towers but could not reliably upload video or reach cloud servers~\cite{Politico2013Wirelesswoes}.

Edge deployment ensures that TeleEMS remains operational in these conditions by keeping all analytics, data exchange, and decision support within the local network. This approach also enables hosting domain-specific large language models, such as EMSLlama for symptom extraction, directly on the edge. Running models locally avoids dependency on commercial cloud LLMs, which may be inaccessible during disasters, produce non-deterministic outputs for identical inputs~\cite{OpenAI2025Platform}, and incur high operational costs. By situating computation at the edge, TeleEMS delivers predictable, cost-effective, and disaster-resilient performance, ensuring uninterrupted decision support during the critical pre-arrival window.

\subsection{EMS-Stream}

\begin{figure}
    \includegraphics[width=\linewidth]{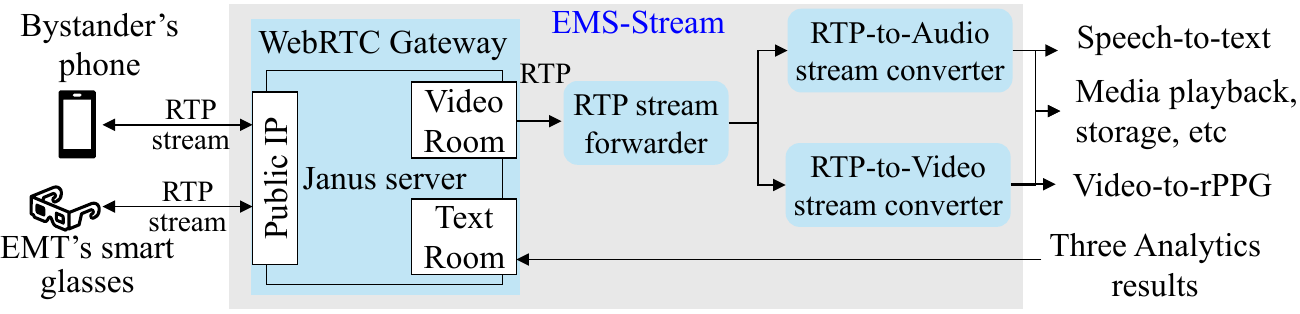}
\caption{EMS-Stream includes a WebRTC Gateway and RTP to video/audio converters. Inside the WebRTC gateway, a Janus server with public local IP and two processes for video room and text room, respectively.}
    \Description{EMS_stream}
    \label{fig:EMS_stream}
\end{figure}

\subsubsection{Design overview}
EMS-Stream is the communication backbone of TeleEMS, enabling low-latency, multi-party transmission of real-time multimedia among bystanders, EMTs, and the TeleEMS server. As shown in Figure~\ref{fig:EMS_stream}, EMS-Stream is implemented as a modular streaming framework composed of a WebRTC Gateway, an RTP stream forwarder, and RTP-to-audio/video converters. These components are designed to integrate seamlessly with the three analytics pipelines in TeleEMS, ensuring that video, audio, and text streams are captured, routed, and processed without breaking the real-time constraints of pre-arrival emergency support.

\subsubsection{Rationale and role in TeleEMS}

The design of EMS-Stream addresses three critical requirements of the TeleEMS system:

\textbf{Multi-party, cross-device interoperability.}
In real-world pre-arrival EMS calls, different participants may join from heterogeneous devices with varying network conditions. WebRTC provides robust NAT traversal, codec negotiation, and cross-platform support, allowing smartphones, smart glasses, and desktop consoles to join seamlessly.

\textbf{Low-latency integration with analytics.}
Pre-arrival intervention guidance must be delivered in seconds, not minutes. By using RTP as the intermediate transport format, EMS-Stream enables direct, zero-copy forwarding of audio and video packets to analytics modules, eliminating unnecessary transcoding or storage delays. This is crucial for feeding real-time streams into EMSLlama for symptom extraction, the video-to-rPPG module for vitals estimation, and PreNet for multimodal task inference.

\textbf{Extensibility for next-generation EMS infrastructure.}
The separation between the WebRTC Gateway, RTP forwarder, and converters allows EMS-Stream to incorporate new analytics modules or replace existing ones without redesigning the streaming core. For example, additional pipelines could be plugged in for bleeding detection, environmental hazard recognition, or automatic triage visualization, all without disrupting the transport layer.

By combining a WebRTC-based gateway with RTP stream forwarding and modular conversion, EMS-Stream ensures that TeleEMS can deliver synchronized, multimodal analytics outputs to all parties in the 911 call session. This design not only enables the current three analytics pipelines but also lays the groundwork for a scalable, open-source communication framework that can evolve with next-generation emergency medical response capabilities.

\subsection{Pre-arrival data preparation for developing analytics modules}
\label{sec:prearrival_data_prep}

TeleEMS is the first system to perform pre-arrival multimodal inference for EMS operations, integrating both symptom texts and patient vitals into a unified decision-making pipeline. TeleEMS trains and evaluates models for three analytics modules: Analytics 1 (audio to patient symptoms), Analytics 2 (video to patient vitals), and Analytics 3 (text-vital to EMS task inference results). Achieving reliable performance for these modules requires large-scale, realistic, pre-arrival datasets. We construct such datasets by transforming the NEMSIS 2023 national EMS database~\cite{nemsis2015,NEMSIS2023Report} into multimodal pre-arrival datasets, supplemented by open-sourced rPPG video and self-collected audio data to fill gaps where no real-world pre-arrival EMS data exists.

\begin{figure}
    \includegraphics[width=\linewidth]{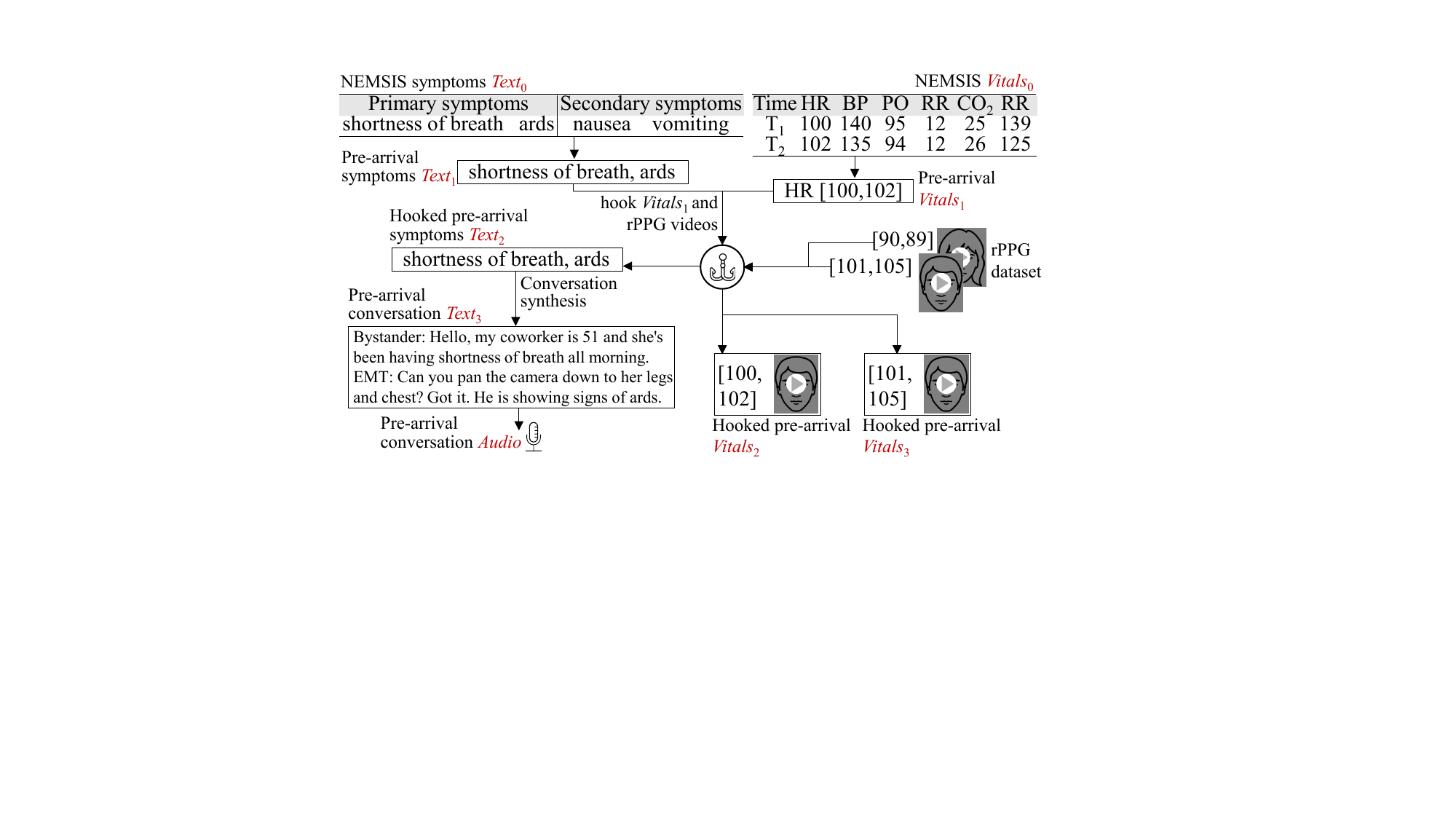}
\caption{Example of using NEMSIS to prepare pre-arrival symptom texts and vitals used in TeleEMS.}
    \Description{symptom_vital_preparation}
    \label{fig:symptom_vital_preparation}
\end{figure}

\subsubsection{Preparing base text and vitals datasets from NEMSIS}

NEMSIS 2023 contains over 50 million EMS incident records in the US, each identified by a unique PcrKey representing a single 911 event. For each incident, we extract symptom texts (primary and secondary symptoms) and time-series vitals (heart rate (HR), blood pressure (BP), pulse oximetry (PO), respiratory rate (RR), end-tidal CO$_2$ (CO$_2$), and blood glucose (BG)), denoted as \textcolor[HTML]{C00000}{\textit{Text}$_0$} and \textcolor[HTML]{C00000}{\textit{Vitals}$_0$}, respectively. An example of \textit{Text}$_0$ and \textit{Vitals}$_0$ are respectively shown on the top of Figure~\ref{fig:symptom_vital_preparation}. After filtering out entries with incomplete symptom and vital fields, we retain approximately 95,000 records, splitting them into train, validation, and test sets in a 3:1:1 ratio. Both \textit{Text}$_0$ and \textit{Vitals}$_0$ have a test set size of 19k, with each \textit{Text}$_0$ sample corresponding to a unique \textit{Vitals}$_0$ sample via PcrKey.

\subsubsection{Extract pre-arrival text and vitals}
\label{sec:design_extract_prearrival_text_vital}

In real 911 calls, primary symptoms (e.g., ``shortness of breath'', ``ards'') are often observable or reportable from streamed bystander videos before EMT arrival, while secondary symptoms (e.g., ``nausea'', ``voitting'') may require on-scene diagnosis and confirmation. We therefore derive pre-arrival symptoms \textcolor[HTML]{C00000}{\textit{Text}$_1$} from \textit{Text}$_0$ by retaining only primary symptoms, resulting same test set size of \textit{Text}$_1$ with that of \textit{Text}$_0$, i.e., 19k, which is shown in Table~\ref{tab:prearrival_motivation_and_rppg_hook}. Similarly, pre-arrival vitals \textcolor[HTML]{C00000}{\textit{Vitals}$_1$} are derived from \textit{Vitals}$_0$ by selecting the HR segment (e.g., HR = [100, 102]) that would plausibly be measurable before arrival, e.g., using remote photoplethysmography (rPPG) from streamed bystander video. This pairing of \textit{Text}$_1$ and \textit{Vitals}$_1$ forms the pre-arrival multimodal dataset for Analytics 3 (PreNet), ensuring that the model trains on a multimodal dataset realistically available before EMT arrival. 

\begin{table}[h]
\caption{Compare MobileBERT's and GRU's performance on pre-arrival and post-arrival text and vital datasets, respectively. TeleEMS uses pre-arrival datasets.}
\label{tab:prearrival_motivation_and_rppg_hook}
\scalebox{0.7}{
\begin{tabular}{llll|cc}
\hline
\multirow{2}{*}{Modality-Model}  & \multicolumn{1}{c}{\multirow{2}{*}{\begin{tabular}[c]{@{}c@{}}Pre/Post\\ arrival\end{tabular}}} & \multicolumn{2}{c|}{Test dataset} & Protocol       & Procedure      \\
                                 & \multicolumn{1}{c}{}                                                                            & Notation          & Size          & Top-1/3/5      & F1 micro/macro \\ \hline
\multirow{4}{*}{Text-MobileBERT} & Post                                                                                            & Text0             & 19k           & 0.79/0.95/0.97 & 0.65/0.33      \\ \cline{2-6} 
                                 & \multirow{3}{*}{Pre}                                                                            & Text1             & 19k           & 0.73/0.88/0.93 & 0.66/0.22      \\
                                 &                                                                                                 & Text2             & 42            & 0.67/0.88/0.95 & 0.70/0.13      \\
                                 &                                                                                                 & Text3             & 42            & 0.60/0.76/0.90 & 0.68/0.17      \\ \hline
\multirow{4}{*}{Vital-GRU}       & Post                                                                                            & Vitals0           & 19k           & 0.49/0.71/0.80 & 0.64/0.13      \\ \cline{2-6} 
                                 & \multirow{3}{*}{Pre}                                                                            & Vitals1           & 19k           & 0.35/0.64/0.73 & 0.61/0.09      \\
                                 &                                                                                                 & Vitals2           & 42            & 0.38/0.74/0.81 & 0.73/0.10      \\
                                 &                                                                                                 & Vitals3           & 42            & 0.38/0.64/0.81 & 0.71/0.10      \\ \hline
\end{tabular}
}
\end{table}

To understand how useful our pre-arrival data is, we use MobileBERT to train, validate, and test on the train set, validate set, and test set of \textit{Text}$_0$ and \textit{Text}$_1$, respectively. MobileBERT is the model used in EMSAssist~\cite{EMSAssist2023MobisysJin,EMSAssistDemo2023} to train for the protocol selection task. As shown in Table~\ref{tab:prearrival_motivation_and_rppg_hook}, the trained MobileBERT on our pre-arrival \textit{Text}$_1$ test set achieves slightly lower top-1, top-3, and top-5 accuracy on the protocol selection task (i.e., Task 1 in Figure~\ref{fig:system_design}, which is basically a multi-class single-label classification task), compared to MobileBERT trained and tested on the post-arrival \textit{Text}$_0$. Despite the small accuracy degradation (from 0.79 to 0.73 top-1 accuracy) due to the inherently less symptom text information in the pre-arrival time window, this result indicates that our preparation of the first multimodal pre-arrival EMS dataset helps retain most of the complete key information in the post-arrival stage.

We also evaluate the two trained MobileBERT models for the Task4 EMT procedure, which is a multiclass multilabel classification task and detailed later in Section~\ref{sec:prenet_design}. We find the pre-arrival \textit{Text}$_1$, without secondary symptoms, enables MobileBERT to achieve very close F1 scores on the procedure recommendation task. For example, pre-arrival \textit{Text}$_1$ achieves 0.66 F1 micro and 0.22 F1 macro scores, while \textit{Text}$_0$ achieves 0.65 F1 micro and 0.23 F1 macro scores. 

Similarly, we train, validate, and test the GRU, a common time-series model, on post-arrival \textit{Vitals}$_0$ and pre-arrival \textit{Vitals}$_1$, respectively. We got the similar observations: pre-arrival \textit{Vitals}$_1$ enables GRU to achieve slightly lower protocol selection and comparable procedure recommendation accuracy, compared to post-arrival \textit{Vitals}$_0$. 

The comparison results between \textit{Text}$_0$ and \textit{Text}$_1$, and \textit{Vitals}$_0$ and \textit{Vitals}$_1$ on post-arrival vs. pre-arrival datasets show slightly lower protocol selection accuracy and comparable performance on the procedure recommendation task. 


\subsubsection{Hooking pre-arrival \textit{Vitals}$_1$ to rPPG videos}
\label{sec:hook_video}


While NEMSIS provides vitals, it contains no patient videos. To enable video-to-vitals evaluation for Analytics 2 and end-to-end evaluation (from video/audio end to EMS tasks end by combining analytics \#1, \#2, and \#3), we generate synthetic but realistic video–vital pairs by matching \textit{Vitals}$_1$ entries with facial videos from the UBFC-rPPG dataset~\cite{UBFCrPPG2019paper,UBFCrPPG2025Dataset}, which contains 42 subject videos with labels of heart rate (HR) time series. As illustrated in Figure~\ref{fig:symptom_vital_preparation}, a video showing a subject’s face along with time-series heart rate (HR) values such as [105, 100] exemplifies the video-HR pairs in UBFC-rPPG. Each video duration is 1 minute, and the paired rPPG and HR labels are both measured within this 1 minute.

Formally, for each heart rate time series \( \mathbf{HR}_i \in \textit{UBFC-rPPG} \), where \( i \in \{1, \dots, 42\} \), we represent it as a 4-tuple:

\begin{equation}
\mathbf{HR}_i = [\mu_i, \sigma_i^2, \gamma_i, \kappa_i]
\label{eq:hr_tuple}
\end{equation}

where \( \mu \) is the mean, \( \sigma^2 \) the variance, \( \gamma \) the skewness, and \( \kappa \) the kurtosis. For each of the 19,000 HR sequences in \textit{Vitals}$_1$, we identify the most similar UBFC-rPPG sequence by minimizing the Euclidean distance:
\begin{equation}
j^* = \arg\min_{j \in \{1, \dots, 19000\}} \; d(\mathbf{HR}_i, \mathbf{HR}_j)
\label{eq:hr_matching}
\end{equation}
where \( d(\cdot, \cdot) \) denotes the Euclidean distance function between two 4-tuples. We mainly use the python scipy package~\cite{2020SciPy-NMeth} to complete this hook process. 

After the hook process, we pair each UBFC-rPPG video with one of the HR sequences in \textit{Vitals}$_1$, these paired HR are \textcolor[HTML]{C00000}{\textit{Vitals}$_2$}. For example, in Figure~\ref{fig:symptom_vital_preparation}, the man face video and corresponding HR [100,105] is paired with [100,102], which is a sample in \textit{Vitals}$_1$. After the hook, we obtain 42 HR samples in \textit{Vitals}$_2$, each corresponding to a video, e.g., [100,102] corresponds to the man face video. At the same time, the original video-HR pair in UBFC-rPPG is also preserved as \textcolor[HTML]{C00000}{\textit{Vitals}$_3$}. It's important to note, as clarified above, a unique sample in \textit{Text}$_1$ corresponds to a unique sample in \textit{Vitals}$_1$. Thus, the hook process will simultaneously get us 42 pre-arrival symptom text samples as \textcolor[HTML]{C00000}{\textit{Text}$_2$}. It's important to note, there are 40 unique pre-arrival symptom text samples among these 42 pre-arrival symptom text samples. This matching process hooks each \textit{Vitals}$_1$ entry with the best-matching UBFC-rPPG video, enabling the construction of synthetic but realistic video-vital pairs for downstream TeleEMS analytics.

Before we use all these vitals data, we apply normally adopted time-series data preprocessing techniques on the HR data, including: removing extremely small or large HR outlier numbers, normalizing all vitals to a decimal between 0 and 1.
It's good to note that the \textit{Text}$_1$ is a subset of the test set of \textit{Text}$_0$. Similarly, \textit{Vitals}$_2$ is a subset of \textit{Vitals}$_1$. So, overall, the hooking process does not influence the model training or validation, it only provides realistic end-to-end pre-arrival test sets for evaluating TeleEMS's performance.

\textbf{Is the hooking process a valid process to prepare pre-arrival video datasets?} The test accuracy related to the hooked \textit{Text}$_2$, \textit{Vitals}$_2$, and \textit{Vitals}$_3$ on protocol selection and procedure recommendation shown in Table~\ref{tab:prearrival_motivation_and_rppg_hook} justifies our hooking process. Although having a degraded 0.60 Top-1 protocol selection accuracy due to the inevitable bias introduced by the smaller data size after the hooking, the hooked \textit{Text}$_2$ achieves 0.88 top-3 and 0.95 top-5, same with \textit{Text}$_1$'s Top-3 at 0.88 and outperform \textit{Text}$_1$'s top-5 at 0.95. On the procedure recommendation, the hooked \textit{Text}$_2$'s F1 micro is larger than \textit{Text}$_1$ by 0.4 while F1 macro lower by 0.9, indicating comparable procedure recommendation accuracy.

On the hooked pre-arrival vitals, we justify by focusing on two comparisons: \textit{Vitals}$_2$ v.s. \textit{Vitals}$_1$ and \textit{Vitals}$_3$ v.s. \textit{Vitals}$_2$. Similar to the comparison between \textit{Text}$_2$ v.s. \textit{Text}$_1$ above, the hooked \textit{Vitals}$_2$ achieves comparable protocol and procedure accuracy with unhooked original NEMSIS HR \textit{Vitals}$_1$. More importantly, the \textit{Vitals}$_3$ achieves the same Top-1, top-5 protocol accuracy and F1 macro accuracy with \textit{Vitals}$_2$. This indicates that using the measured HRs labels of UBFC-rPPG video dataset has the nearly the same effect as directly using the NEMSIS HRs, which is the essential observation on which that our hooking justification is based, and also what the hooking Equation~\ref{eq:hr_matching} indicates.

Thus, this two comparisions (text and vitals) justify the validity of the hooking process to prepare the pre-arrival video dataset.

\subsubsection{Conversation synthesis and audio recording} 
\label{sec:conversation_and_audio_prep}

TeleEMS aims to build an end-to-end (E2E) system to advance pre-arrival EMS delivery—from streamed bystander videos during 911 calls to EMS task outputs. Next-generation 911 calls include both videos and audio streamed from callers to EMTs. Section~\ref{sec:hook_video} resolves the video part through the hook process, but no existing E2E dataset contains complete real-world EMT-bystander conversations of EMS events. Although NEMSIS has EMS symptoms \textit{Text}$_0$, we get pre-arrival symptoms \textit{Text}$_1$ from \textit{Text}$_0$, we get \textit{Text}$_2$ from the hook process, but no existing E2E dataset contains complete real-world EMT-bystander conversations of EMS events. To address this limitation, for each \textit{Text}$_2$ sample, we synthesize a conversation between EMTs and bystanders by augmenting the two pre-arrival symptoms in the \textit{Text}$_2$ sample with non-symptom expressions. These non-symptom expressions are commonly seen and used by EMTs and callers in daily 911 calls. 42 such synthesized conversations would be the pre-arrival conversation \textcolor[HTML]{C00000}{\textit{Text}$_3$}, 40 of which are unique. 

Given 40 unique synthesized \textit{Text}$_3$, we recruit four volunteers to speak these 40 conversations loudly. We record their voice of these pre-arrival conversation using three microphones: PH1 phone~\cite{PH1Specs2025}, Google Glass Enterprise II~\cite{GoogleEEII2024,glassSpecs2023}, and professional microphone HyperX~\cite{HyperXSolocast2024}. In total, we have $40 * 4 * 3 = 480$ pre-arrival conversation \textcolor[HTML]{C00000}{\textit{Audio}} dataset.

\subsection{Analytics \#1: Speech-to-text and EMSLlama}

\begin{figure}
    \includegraphics[width=\linewidth]{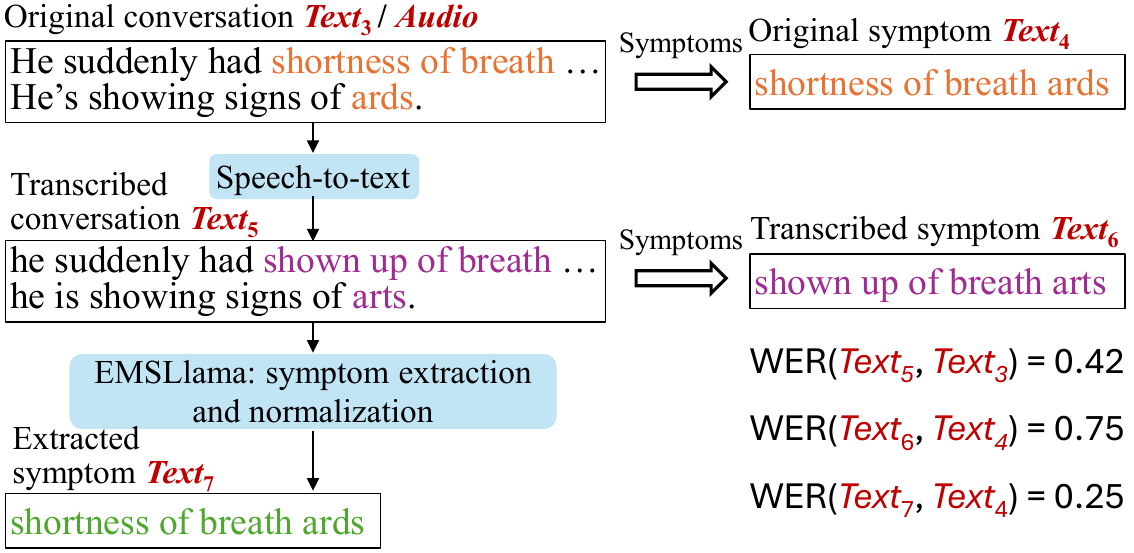}
    \caption{Speech-to-text and symptom extractor pipeline in TeleEMS (left) and corresponding symptoms in the EMT-bystander conversation (right).}
    \Description{sr_medner_pipeline}
    \label{fig:sr_medner_pipeline}
\end{figure}

Figure~\ref{fig:sr_medner_pipeline} illustrates the speech-to-text and EMSLlama pipeline details. Initially, the EMT-bystander conversation audio example, \textit{Text}$_3$ or \textit{Audio} are obtained from Section~\ref{sec:conversation_and_audio_prep} discussed above, contains two symptoms (\textcolor[HTML]{C00000}{\textit{Text}$_4$}): ``shortness of breath'' and ``ards''. A speech-to-text model (e.g., Google Cloud default API~\cite{google_speechtotext_models2025}), although correctly transcribing the non-symptom texts (\textcolor[HTML]{C00000}{\textit{Text}$_5$}), incorrectly transcribes two symptoms as ``shown up of breath'' and ``arts'' (\textcolor[HTML]{C00000}{\textit{Text}$_6$}), respectively. After that, our EMSLlama, a fine-tuned Llama-3-8b model~\cite{llama8bhugginface,grattafiori2024llama3herdmodels} takes the entire transcribed conversation \textit{Text}$_5$ as input to extract and normalize the two incorrectly transcribed symptoms into ``shortness of breath'' and "ards" (\textit{Text}$_7$), which are the same with the original symptoms \textit{Text}$_4$. 

\subsubsection{Speech-to-text} 
\label{sec:design_speech_to_text}
In TeleEMS, we adopt the SOTA open-sourced speech-to-text whisper series models from OpenAI, which is also used in the CognitiveEMS~\cite{CognitiveEMS2024IoTDI}, the SOTA post-arrival voice assistant for EMS. We evaluate whisper model's performance on the \textit{Audio} dataset described in Section~\ref{sec:conversation_and_audio_prep}. We compare whisper with other speech-to-text baselines: EMSConformer, Vosk series, and Google Cloud APIs. The details of whisper and baselines are as follows:

\begin{itemize}
    \item EMSConformer in EMSAssist is a Conformer model~\cite{conformer2020,conformer_github2022} fine-tuned on their collected audio dataset of only EMS patient symptoms~\cite{EMSAssist2023MobisysJin}.
    \item Four Alpha Cephei Vosk models are Kaldi-based~\cite{kaldi_ASRU2011} speech-to-text models fully developed on general audio datasets~\cite{librispeech2015,tedlium2012}: vosk-small (vosk-model-small-en-us-0.15), vosk-ml (vosk-model-en-us-0.22-lgraph), vosk-medium (vosk-model-en-us-0.22), and vosk-large (vosk-model-en-us-0.42-gigaspeech)~\cite{alphacephei,VoskModels}.
    \item Three Google Cloud Speech-to-Text APIs: GC-default, GC-med (medical conversation), and GC-video~\cite{google_speechtotext_models2025}.
    \item Four OpenAI Whisper models~\cite{whisper2023icml}: whisper-tiny, whisper-base, whisper-small, and whisper-medium.
\end{itemize}

\subsubsection{EMSLlama motivation: Fine-tuned Llama for symptom extraction and normalization}

Before we introduce the design of EMSLlama, we first explain the motivations for 1) symptom extraction and normalization and 2) for using EMSLlama to do symptom extraction and normalization.

\textbf{Why do we need symptom extraction and normalization in TeleEMS?} We mainly consider the following two motivations:

\textbf{Motivation 1}: When compared to conversation texts, no matter transcribed or original, symptom texts increase protocol selection accuracy. In Table~\ref{tab:prearrival_motivation_and_rppg_hook}, the \textit{Text}$_3$ is a text dataset of 42 EMT-bystander conversations, each containing two primary symptoms, i.e., \textit{Text}$_2$. The test results show that \textit{Text}$_3$ has consistently and obviously lower accuracy on the protocol selection task. This is because the conversation texts, although containing primary symptoms texts, major portion of which is the non-symptom texts. These non-symptom texts, if not filtered out, will degrade the accuracy. So, extracting the symptoms from conversations can increase the accuracy. 

\textbf{Motivation 2}: Symptom extraction and normalization align with the goal of next generation EMS infrastructure in retaining the key symptom in a 911 call session. Although the current-generation infrastructures provide live audio transcription services, there are inevitable transcription errors, especially when the speaker is in a high-stress and noisy environment. With symptoms extracted and normalized into a standard form that EMTs are familiar with, EMTs don't need to spend time tracking the mentions of these symptoms or verifying the ground truth of incorrectly transcribed symptoms. 

\textbf{Why do we need to use EMSLlama for symptom extraction and normalization in TeleEMS?}


\textbf{Motivation 3}: The emerging large language model shows high potential in various NLP tasks, including key clinical information retrieval~\cite{LLMRag2025NPJ,LLMMedicalEncode2023nature,LLMClinicalTextAnalysis2024npj,hu2025harnessing}. However, most of them focus on retrieving clinical information from well-written medical texts. They only need to extract key symptoms from well written medical texts, without need to worry about the normalization. In TeleEMS, the key symptoms are usually embedded inside the conversation audios and may be incorrectly transcribed. The requirement of accurately extracting incorrectly transcribed symptoms from the conversation transcriptions make those solutions inapplicable.

\begin{figure*}[t]
    \centering
    \begin{subfigure}[t]{0.64\linewidth}
        \includegraphics[width=\linewidth]{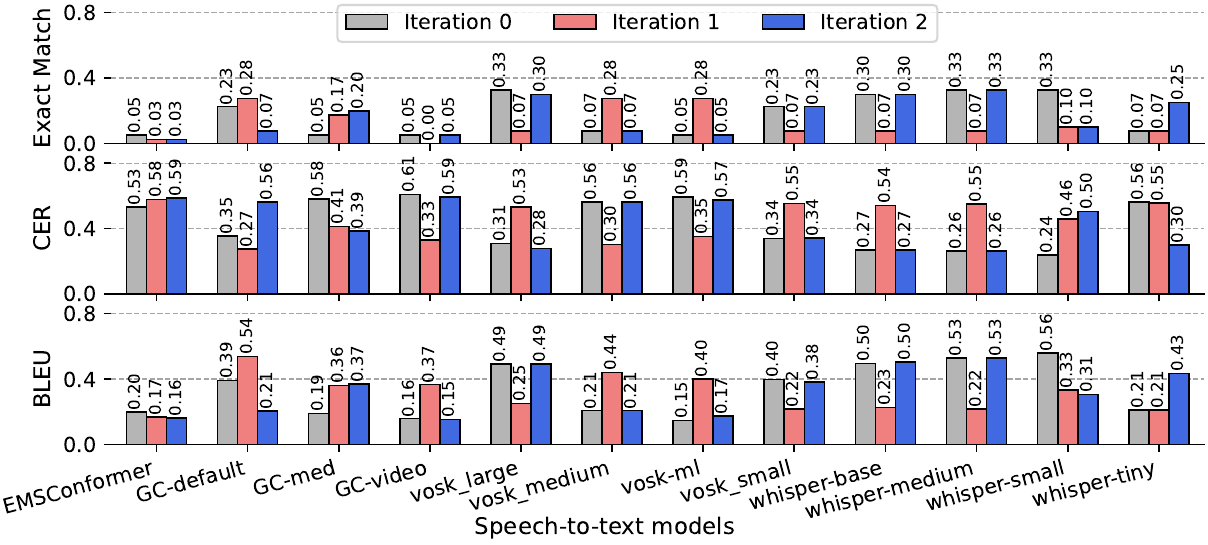}
        \caption{GPT4o's high indeterminism in the symptom extraction and normalization task. 
        In each of the three iterations, we feed the same input conversation transcription prompt, GPT4o outputs rapidly changing symptom outputs.}
        \label{fig:gpt4o_indeterminism}
    \end{subfigure}
    \hfill
    \begin{subfigure}[t]{0.34\linewidth}
        \includegraphics[width=\linewidth]{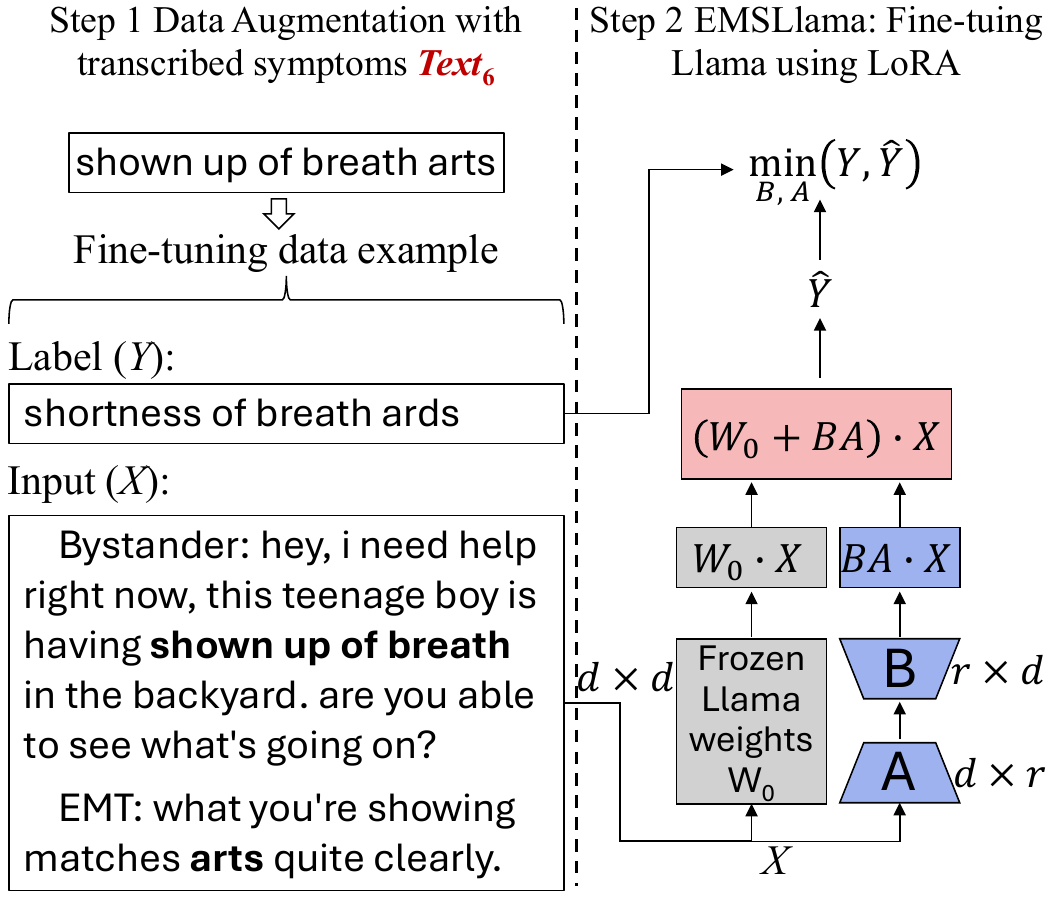}
        \caption{Two-step design of EMSLlama. Step 1: data augmentation with transcribed symptoms; step 2: fine-tuning Llama with the augmented dataset (\textit{X}-\textit{Y} pairs) using low-rank adaptation (LoRA).}
        \label{fig:EMSLlama}
    \end{subfigure}
    \caption{(a) GPT4o's high indeterminism in symptom extraction and normalization. (b) Two-step EMSLlama design.}
    \Description{Combined GPT4o indeterminism and EMSLlama design.}
    \label{fig:combined}
\end{figure*}

Although commercial LLMs (e.g., GPT-4o) are trained on various texts, showing the potential of extracting and normalizing the symptoms from the conversation transcriptions, their outputs are neither accurate nor deterministic, i.e., their outputs are not accurate and are different given the same inputs. For example, OpenAI specifically says the determinism is not guaranteed when prompting their platform APIs~\cite{OpenAI2025Platform} with same input prompts. Figure~\ref{fig:gpt4o_indeterminism} corroborates such claims. 
For example, for the speech-to-text model ``whisper-tiny'', we apply the whisper-tiny model to transcribe audio recorded by a single user using a single microphone. We use the GPT-4o to extract and normalize the symptoms in these whisper-tiny transcriptions. We then compare the extracted and normalized symptoms with the true symptoms. Figure~\ref{fig:gpt4o_indeterminism} illustrates the average comparison results under three metrics commonly used for evaluting the similarities between two strings: exact match for evaluating if the two strings are identical, the higher the more performant of the symptom extraction and normalization model; character error rates (CER) for evaluating the character level descrepancy between two strings, the lower the better; bilingual evaluation understudy (BLEU) for evaluating the overall text quality aligning with human judgement, a metric commonly used in machine translation tasks, the higher the better. We input the same prompts, including the instructions and conversation transcriptions, as shown in Appendix~\ref{sec:input_prompt_gpt4o}.

As shown, overall, the GPT-4o does not perform well on all three metrics across all speech-to-text models, including Google Cloud APIs and whisper series models. For example, the exact match rates are all below 0.4, most CER are above 0.5, and the maximum BLEU is capped at 0.55 across all speech-to-text models. The highly variable results across the three iterations corroborates that GPT-4o's outputs are rapidly changing, even with the same prompt inputs. The low extraction and normalization accuracy and high indeterminism make the commerical LLMs like GPT-4o unapplicable for life-threatending task like pre-arrival EMS.

\subsubsection{EMSLlama design: Fine-tuning Llama-3-8B with low-rank adaptation}

To enable acurate and robust symptom extraction and normalization in noisy, pre-arrival EMS scenarios, we fine-tune the Llama-3-8B model~\cite{llama8bhugginface,grattafiori2024llama3herdmodels} using low-rank adaptation (LoRA)~\cite{hu2021lora}. Direct full-parameter fine-tuning of Llama-3-8B is computationally prohibitive, requiring hundreds of gigabytes of GPU memory and long training cycles. LoRA addresses this limitation by freezing the original model weights $\mathbf{W}_0$ and introducing two small trainable low-rank matrices $\mathbf{A}$ and $\mathbf{B}$ into each targeted layer. The adapted weight is then expressed as:
\[
\mathbf{W} = \mathbf{W}_0 + \mathbf{B}\mathbf{A}.
\]
Because $\mathbf{A}$ and $\mathbf{B}$ are of rank $r \ll d$, the number of trainable parameters is reduced by orders of magnitude, yet the model retains sufficient expressivity for domain-specific adaptation. Our EMSLlama pipeline consists of two stages, illustrated in Figure~\ref{fig:EMSLlama}:  

\textbf{Step~1: Data augmentation.} We first generate large-scale $(\textit{X}, \textit{Y})$ pairs from raw transcribed symptoms. From 480 noisy transcriptions, we manually extract the symptom phrases and inject each into a unique EMT–bystander conversation template. This simulates realistic speech-to-text (STT) conditions where non-symptom portions of the conversation vary, but the injected symptom reflects true STT symptom outputs. Here, $\textit{X}$ is the conversation with an injected transcribed symptom, and $\textit{Y}$ is the corresponding correctly normalized symptoms. The dataset is split into 384 training pairs and 96 validation pairs (4:1 ratio). For the test pairs, we use transcribed conversations \textit{Text}$_5$ exemplified in Figure~\ref{fig:sr_medner_pipeline} as the input, and corresponding truth symptoms \textit{Text}$_4$ as the output. The size of the test set is equal to (number of speech-to-model) * (number of microphone type) * (number of audio participants) * (number of audios recorded by a participant) = 12 * 3 * 4 * 40 = 5760.

Importantly, the non-symptom text in $\textit{X}$ is normalized rather than transcribed verbatim, so the gold evaluation in Section~\ref{sec:evaluate_llama} is performed on actual conversation transcriptions to measure real-world correctness and robustness. This augmentation strategy compensates for the scarcity of labeled pre-arrival EMS STT data, particularly for rare or colloquial symptom descriptions.

\textbf{Step~2: LoRA fine-tuning.} We fine-tune Llama-3-8B using LoRA with a rank $r=16$ in all adapted layers. Only $\mathbf{A}$ and $\mathbf{B}$ are updated during fine-tuning; $\mathbf{W}_0$ remains frozen. This avoids forgetting of general language capabilities while enabling targeted adaptation to EMS-specific terminology and context. The fine-tuned $\mathbf{A}$ and $\mathbf{B}$ matrices are small enough to be stored separately and merged into $\mathbf{W}_0$ at inference time. In deployment, EMSLlama loads the frozen $\mathbf{W}_0$ from the original Llama-3-8B checkpoint and overlays the learned $\mathbf{A}$ and $\mathbf{B}$ matrices on top. This allows the full model to be reconstructed for inference without retraining, while only needing to keep the matrices $A$ and $B$ on deployable storage, which is critical for edge or on-premise EMS dispatch servers.

EMSLlama’s novelty lies in three aspects: (\textit{i}) operationalizing Llama-3-8B for a safety-critical, latency-sensitive EMS setting; (\textit{ii}) leveraging a targeted data augmentation strategy grounded in real-world STT symptom outputs; and (\textit{iii}) achieving domain-specific adaptation under strict memory and compute constraints, enabling practical deployment in the TeleEMS system.

\subsection{Analytics \#2: Video-to-rPPG and rPPG-to-Vitals}
\label{sec:design_analytics2_video_to_vitals}

As illustrated in Figure~\ref{fig:system_design} and detailed in Section~\ref{sec:prearrival_data_prep}, the second analytics module in TeleEMS estimates patient vital signs from live facial video streams. This is achieved by applying SOTA remote photoplethysmography (rPPG) methods, which recover physiological signals from subtle skin color variations caused by blood volume changes. The incoming video stream is segmented into consecutive, non-overlapping six-second windows, which provide a balance between capturing rapid physiological changes and maintaining computational efficiency.

For each video segment, we apply the TSCAN model~\cite{rppgtoolbox2023neurips,tscan2020neurips} to extract the rPPG waveform. This waveform is a time series that reflects the minute variations in skin reflectance induced by the cardiac pulse. The rPPG signal then passes through an open-source vital sign extraction pipeline~\cite{rPPG2HRViz}. The pipeline first removes long-term trends in the signal with a smoothing filter, then applies a band-pass filter to retain only the frequency range between 0.75 and 2.5 hertz. This range corresponds to physiologically plausible heart rates (HR) between 45 and 150 beats per minute. The dominant frequency, determined by the peak in the power spectral density, is taken as the HR for the corresponding window. As the process repeats, we can get a continuous time series of HR estimates from live facial video, enabling real-time tracking of patient cardiovascular status during pre-arrival emergency care.

The broader research community has shown increasing interest in contactless vital sign monitoring. In particular, rPPG techniques have matured from early laboratory demonstrations to prototypes deployed in consumer devices such as smartphones, and even to commercial products~\cite{FacePhys2025,Biosensing2025}. In this work, TeleEMS does not attempt to improve rPPG algorithms themselves for two key reasons. First, while academic datasets such as UBFC-rPPG exist, real-world rPPG video datasets—especially those recorded in high-stress emergency conditions—are not yet publicly available. Without such datasets, it is not feasible to rigorously evaluate improved rPPG algorithms or to adapt state-of-the-art rPPG models to the conditions encountered during emergency medical services. Second, our focus is on demonstrating the integration of an rPPG module into a complete, streaming video analytics pipeline for pre-arrival EMS care. TeleEMS’s modular architecture is explicitly designed to support plug-and-play replacement of the rPPG component, allowing any research prototype, open-source library, or commercial-grade model to be seamlessly swapped in as new capabilities become available. This design choice ensures that TeleEMS remains future-proof while delivering immediate value as a unified, deployable system.


\subsection{Analytics \#3: PreNet}
\label{sec:prenet_design}

\begin{figure}
    \includegraphics[width=\linewidth]{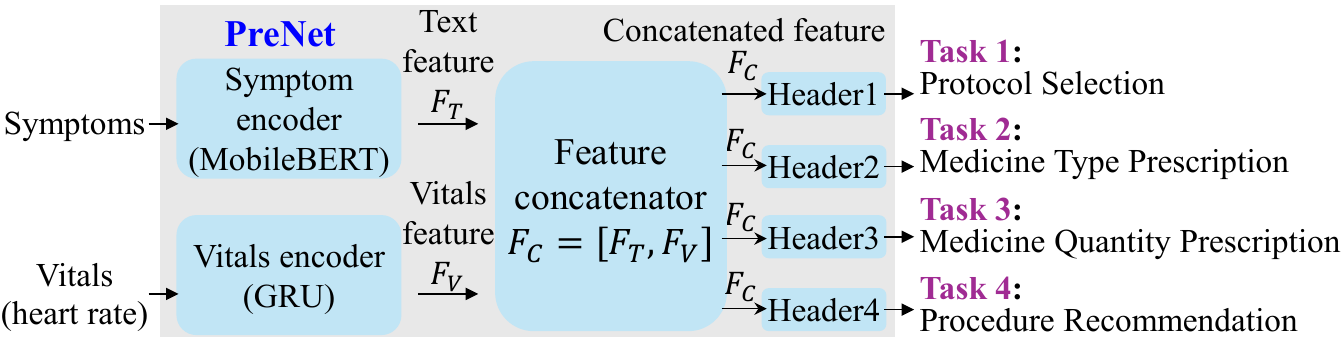}
\caption{Design of PreNet.}
    \Description{PreNet}
    \label{fig:prenet}
\end{figure}

As illustrated in Figure~\ref{fig:prenet}, PreNet is a multimodal multitask network purpose-built for TeleEMS to integrate symptom texts from EMSLlama and vital signs from rPPG-based video analysis into a unified decision-making model. The architecture begins with two modality-specific encoders: the symptom encoder (MobileBERT) transforms textual symptom descriptions into a text feature vector $F_T$, while the vitals encoder (GRU) processes numerical vital sign time series, such as heart rate, into a vitals feature vector $F_V$. These representations are concatenated to form a joint multimodal embedding $F_C = [F_T, F_V]$, ensuring that both verbal and physiological cues are preserved in a shared feature space.

From this shared feature space, $F_C$ is passed to four specialized output heads, each aligned with a specific EMS decision task. Header1 predicts the emergency medical protocol, Header2 determines the medication type, Header3 estimates the appropriate medication quantity, and Header4 recommends relevant medical procedures. This design allows correlations between modalities and tasks to be exploited, e.g., a high heart rate (from $F_V$) combined with specific breathing-related symptom terms (from $F_T$) can reinforce the selection of a protocol in Header1 and influence dosage estimation in Header3.

This design is tailored for the TeleEMS system in three key ways. First, it integrates multimodal pre-arrival data (verbal symptom texts and physiological vitals) into a single predictive pipeline. Second, it enables real-time inference by using lightweight encoders and shared representations, which reduces computational overhead. Third, due to PreNet's modular design, it is flexible enough to accommodate additional input modalities or decision-making tasks in future EMS deployments, making it well suited for evolving field requirements.

\section{TeleEMS Implementation}
\label{sec:implementation}

The TeleEMS server is implemented in approximately 3,000 lines of Python code and deployed on an edge server equipped with three NVIDIA A30 GPUs within a campus network. The TeleEMS client consists of about 2,000 lines of Android Java code for deployment on two platforms: an Android smartphone (used by the bystander) and a Google Glass Enterprise II smart glass (used by an EMT en route). In addition, a desktop client for the 911 dispatcher is implemented in roughly 500 lines of Python code. We used the Janus Server~\cite{JanusGithub2025,IPTComm2014-Janus} as the backbone of the EMS-Stream. After deployment, EMS-Stream runs as a daemon process, and all analytics run in a separate process on the edge server.

All training, validation, and testing of the three analytics pipelines are conducted using PyTorch on the same edge server used for deployment. For model training, we adopt widely used hyper-parameter configurations rather than exhaustively tuning them. For example, when fine-tuning EMSLlama for Analytics\#1, we fix the LoRA rank parameter at $r=16$, and when training PreNet for Analytics\#3, we use a fixed learning rate of $5\times10^{-5}$. This approach prioritizes demonstrating the feasibility and integration of the analytics modules within the end-to-end TeleEMS pipeline over exhaustive hyper-parameter optimization.

\section{TeleEMS Evaluation}
\label{sec:evaluation}


\subsection{Speech-to-text transcription}

In Figure~\ref{fig:sr_medner_pipeline} and described in Section~\ref{sec:conversation_and_audio_prep}, we obtained an audio dataset of size 480, which is the first pre-arrival audio dataset that reflects the real-world pre-arrival EMT-bystander conversations. We apply 12 speech-to-text models listed in Section~\ref{sec:design_speech_to_text}. The 12 models cover the SOTA speech-to-text models from industry (e.g., Google Cloud APIs~\cite{google_speechtotext_models2025,google_speechtotext2025}), from prior EMS research work (e.g., EMSConformer used in EMSAssist~\cite{EMSAssist2023MobisysJin,EMSAssistDemo2023}), and from open-sourced community (e.g., whisper series~\cite{whisper2023icml} and vosk series~\cite{VoskModels}), representing the best effort in speech-to-text on transcribing the pre-arrival EMT-bystander audios.

\begin{figure}[h]
    \includegraphics[width=\linewidth]{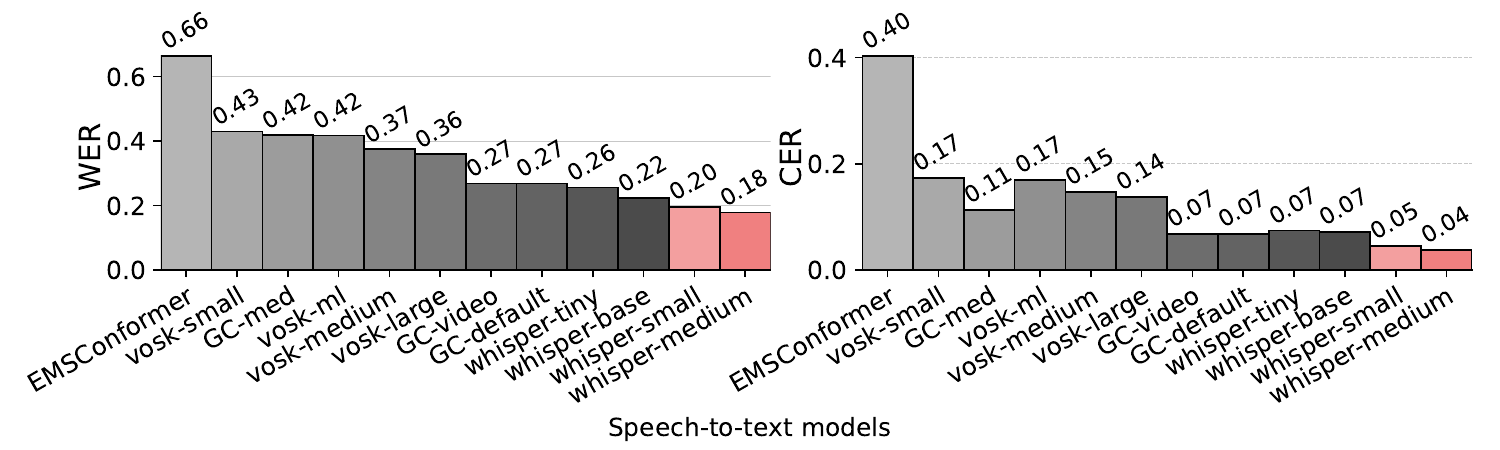}
    \caption{Average word error rate (WER) and character error rate (CER) of different speech-to-text transcription models on our collected conversation audio dataset, descendingly sorted by the WER. The lower WER and CER, the better the model performance.}
    \Description{sr_model_wer_cer}
    \label{fig:sr_model_wer_cer}
\end{figure}

Figure~\ref{fig:sr_model_wer_cer} illustrates the performance of these 12 models using the word error rates (WER) and character error rates (CER) as the metrics, the lower the more accurate transcriptions. To better identify the best speech-to-text models, i.e., whisper-small and whisper-medium, We sorted the 12 speech-to-text models with descending WERs.

\begin{figure}
    \includegraphics[width=\linewidth]{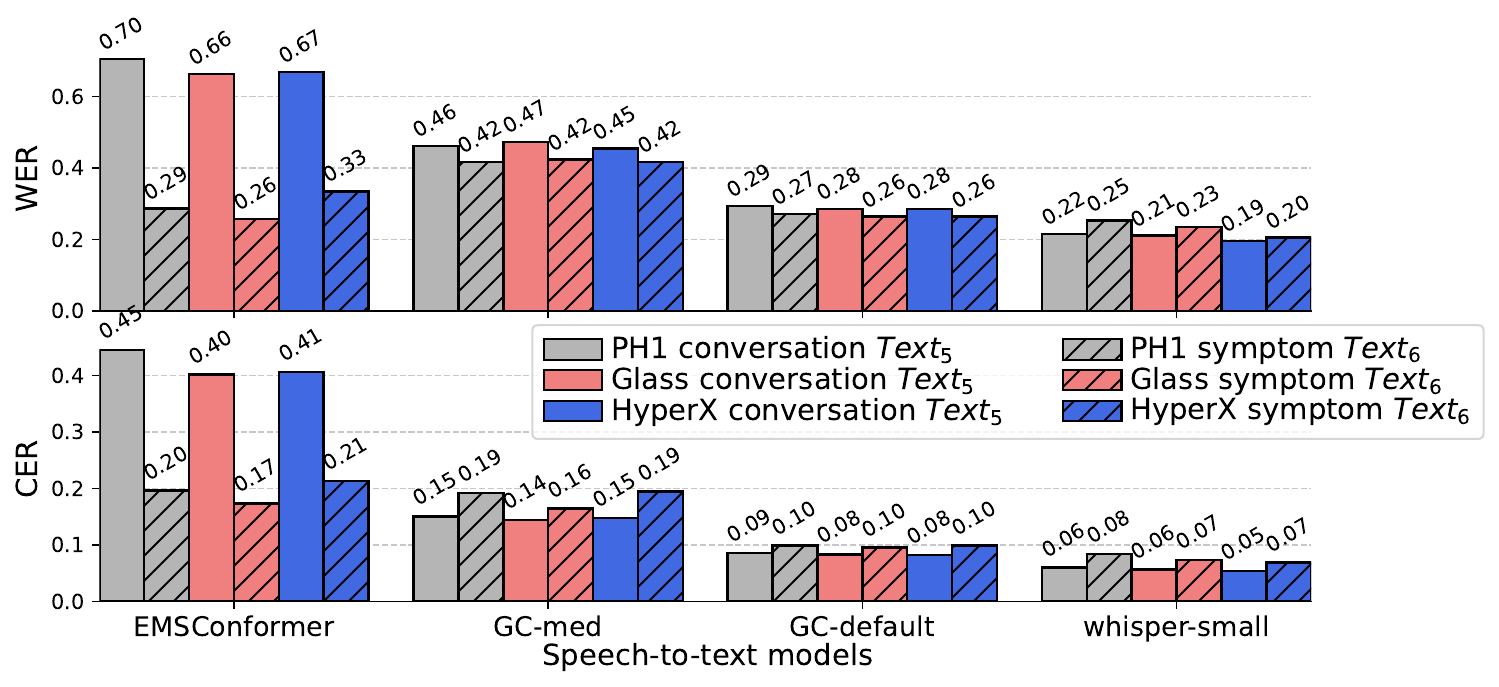}
    \caption{Speech-to-text model performance on a user's conversation audio dataset: $Text_3$ vs. $Text_1$ and $Text_4$ vs. $Text_2$.}
    \Description{conversation_manual_bioner_wer_cer_cropped}
    \label{fig:conversation_manual_bioner_wer_cer}
\end{figure}

Figure~\ref{fig:conversation_manual_bioner_wer_cer} reveals that word error rates (WER) and character error rates (CER) of full conversations do not always reflect the accuracy of extracted symptom descriptions. Across different speech-to-text models and microphones, the gap between solid (conversation-level) and shaded (symptom-level) bars varies significantly. In some cases (e.g., EMSAssist or GC-med), symptom WER is substantially lower than full conversation WER, suggesting that even noisy transcripts may retain clinically important phrases. In other cases, symptom-level errors remain high despite low overall WER.

These inconsistencies highlight that evaluating only full-conversation transcription quality is insufficient. A dedicated evaluation of extracted symptoms is necessary to ensure clinical relevance. This motivates our approach of explicitly extracting and assessing symptoms from transcribed conversations as a standalone task. 

Overall, whisper-medium and whisper-small achieve the lower WER and CER than all other speech-to-text models, including the three Google Cloud APIs. We mainly use whisper-medium and whisper-small as the speech-to-text task component in TeleEMS.

\subsection{Performance of EMSLlama}
\label{sec:evaluate_llama}

\begin{figure*}
    \includegraphics[width=\linewidth]{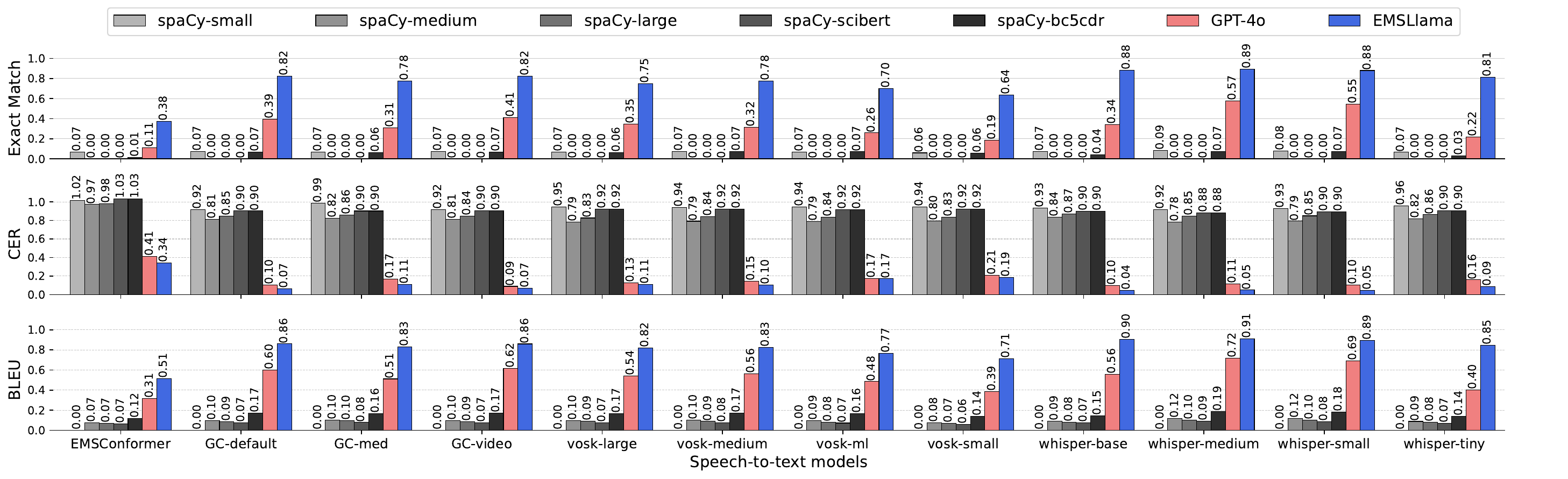}
    \caption{Comparing transcribed symptoms extracted by different MedNER models with original untranscribed symptom texts.}
    \Description{medner_truth.}
    \label{fig:bioner_truth}
\end{figure*}

We evaluate EMSLlama against GPT-4o and traditional biomedical NER baselines to assess its ability to extract and normalize symptoms from noisy transcriptions.
Figure~\ref{fig:bioner_truth} summarizes results across three metrics including exact match, character error rate (CER), and BLEU. These metrics comparing model outputs against ground-truth symptoms, i.e., \textit{Text}$_7$ v.s. \textit{Text}$_4$ as shown in Figure~\ref{fig:sr_medner_pipeline}. We benchmark five widely used spaCy-based biomedical NER models~\cite{scispacy2019medner}, including small, medium, large, scibert (fine-tuned from scibert-base~\cite{Beltagy2019SciBERT}), and bc5cdr (trained on the BC5CDR biomedical chemical corpus~\cite{BC5CDR2016}). We also evaluate GPT-4o, noting that its outputs vary across runs even with identical prompts (Figure~\ref{fig:gpt4o_indeterminism}). To ensure fairness, we report GPT-4o’s best results over three repeated runs. All models are evaluated using transcriptions from the same 12 speech-to-text systems introduced in Figure~\ref{fig:gpt4o_indeterminism} and Figure~\ref{fig:sr_model_wer_cer}, with consistent evaluation metrics.


\textbf{EMSLlama achieves consistently superior performance, delivering more accurate and reliable symptom normalization.}
Across all speech-to-text inputs, EMSLlama produces the highest exact match scores, the lowest CER, and the highest BLEU values, outperforming both spaCy baselines and GPT-4o. For example, using transcriptions from whisper-medium, GPT-4o achieves an exact match rate of 0.57, while EMSLlama raises this to 0.89. This improvement of more than 30 percentage points highlights the strength of domain-specific fine-tuning for symptom extraction. Importantly, EMSLlama’s outputs are deterministic and reproducible, whereas GPT-4o exhibits non-determinism that complicates downstream integration.


\textbf{EMSLlama's practical implications for pre-arrival EMS support are significant.}
As discussed in Section~\ref{sec:challenge_contribution}, EMTs require clean and normalized symptoms rather than verbatim noisy transcripts to prepare interventions and guide bystanders in real time. By consistently outperforming GPT-4o’s strongest outputs and eliminating variability across runs, EMSLlama provides a more dependable foundation for pre-arrival decision support. This advantage makes EMSLlama better suited for real-world deployment in TeleEMS, where accurate symptom normalization directly translates to more actionable clinical interventions.

\subsection{End-to-end evaluation}

We now evaluate the end-to-end performance of the TeleEMS analytics pipeline. By end-to-end, we mean the complete path from the audio and video forwarded through EMS-Stream to the final outputs of the four EMS inference tasks: protocol selection, medicine type prescription, medicine quantity prescription, and procedure recommendation. This evaluation directly reflects the system’s ability to deliver actionable insights from pre-arrival data.

Following the design in Section~\ref{sec:design_extract_prearrival_text_vital} and the metrics used in Table~\ref{tab:prearrival_motivation_and_rppg_hook}, we assess task 1 (protocol selection) and task 2 (medicine type prescription) with top-1, top-3, and top-5 accuracy. For task 4 (procedure recommendation), which is a multi-label classification problem, we use F1 micro and F1 macro. For task 3 (medicine quantity prescription), a single-value regression, we use mean squared error (MSE), Pearson coefficient~\cite{Pearsonr}, and Spearman coefficient~\cite{Spearman}. For all metrics except MSE, higher values indicate better performance. Since PreNet is a multimodal and multitask model, we consider both single-input versus multi-input scenarios and single-task versus four-task simultaneous inference.

\subsubsection{End-to-end text evaluation}
\label{sec:eval_e2e_text_eval}

We first evaluate the text-only pipeline, where vitals are not included, and measure accuracy when audio transcriptions are processed through analytics \#1 and \#3. For symptom extraction and normalization, we compare four options: (i) None, where raw conversation texts are used without symptom extraction, including both ground-truth conversations and speech-to-text transcriptions; (ii) Manual, where normalized symptoms are obtained from human annotators; (iii) GPT-4o, where symptoms are extracted and normalized using the prompt shown in Appendix~\ref{sec:input_prompt_gpt4o}; and (iv) EMSLlama, where symptoms are extracted and normalized by our fine-tuned EMSLlama model. As Figures~\ref{fig:sr_model_wer_cer} and~\ref{fig:conversation_manual_bioner_wer_cer} showed earlier, the Whisper series and Google Cloud APIs provide the highest transcription accuracy, so their transcriptions form the inputs of our evaluations.

First, across all speech-to-text models, accuracies for tasks 1, 2, and 3 are significantly higher when symptom extraction and normalization is included (Manual, GPT-4o, or EMSLlama) compared to the None baseline. For example, on the top-3 protocol selection single task, EMSLlama achieves 0.89 accuracy with Whisper-small transcriptions, very close to the 0.90 accuracy of human-annotated symptom truth. Without symptom extraction, the best accuracy observed is 0.75 using Google Cloud transcriptions. This result validates our decision to integrate an explicit symptom extraction and normalization stage into analytics \#1.

Second, EMSLlama delivers performance comparable to GPT-4o across all tasks, all speech-to-text models, and both single-task and four-task settings. On more than half of the evaluation columns, EMSLlama achieves the highest accuracy, while GPT-4o leads in the remaining cases. Even in those cases, EMSLlama remains very close. For instance, when PreNet simultaneously executes all four tasks, EMSLlama enables a top-3 medicine type prescription accuracy of 0.93 with Google Cloud transcriptions, higher than GPT-4o's 0.89–0.90. Similarly, although GPT-4o achieves the best top-1 medicine type prescription accuracy of 0.57, EMSLlama achieves 0.56 using Whisper-tiny transcriptions, essentially matching GPT-4o despite being an open fine-tuned model.

Third, the procedure recommendation task demonstrates how pre-arrival analytics directly support intervention decision-making. Here, our model achieves an F1 micro of 0.68–0.69 and an F1 macro of 0.12–0.20. The relatively high F1 micro indicates that the system reliably predicts frequent and critical interventions that dominate EMS operations. The lower F1 macro reflects the inherent difficulty of predicting rare interventions that appear only in a small fraction of cases. Importantly, the strong F1 micro suggests that TeleEMS can already provide actionable decision support for common, high-impact interventions such as oxygen administration, medication delivery, and transport decisions, which represent the majority of emergency encounters. At the same time, the gap in macro F1 highlights a meaningful research challenge: handling label imbalance to better capture rare but critical interventions.

To better under the evaluation results in the procedure recommendation, the achieved F1 micro of 0.68–0.69 is consistent with the 0.5–0.7 F1 micro reported by BERT-based models trained on discharge notes in the MIMIC-III dataset~\cite{heo2021medicalcodepredictiondischarge,PONTHONGMAK2023101227,Huang_2019}. Likewise, the F1 macro range of 0.12–0.20 aligns with prior work on multi-label classification in medical texts, where even state-of-the-art models yield macro F1 in the 0.04–0.15 range on extreme label distributions~\cite{nguyen2023mimicivicdnewbenchmarkextreme}. More recent fine-tuned LLMs on clinical notes report macro F1 between 0.2 and 0.6, underscoring how heavily imbalanced medical label distributions suppress macro performance even in well-studied settings. Since both our prepared procedure dataset and the underlying NEMSIS source are highly unbalanced, the observed performance is expected and provides a foundation for further improvement in actionable rare-case recommendations.

\begin{table*}[]
\caption{End-to-end EMS task accuracy using text-only data (audio-only).}
\label{tab:e2e_text_eval}
\scalebox{0.7}{
\begin{tabular}{llcccccccc}
\hline
Symptom       & Speech-         & \multicolumn{4}{c}{Single task}                                  & \multicolumn{4}{c}{Multiple task (4 simultaneous tasks)}          \\
extraction    & to-text    & Protocol       & MedType        & Quantity       & Procedure(F1) & Protocol       & MedType        & Quantity        & Procedure(F1) \\
normalization & model          & Top1/3/5       & Top1/3/5       & MSE/Pear/Spear & micro/macro   & Top1/3/5       & Top1/3/5       & MSE/Pear/Spear  & micro/macro   \\ \hline
\multirow{8}{*}{None} &
  conversation truth &
  0.60/0.76/0.90 &
  0.45/0.88/0.93 &
  7.67/0.27/0.24 &
  0.68/0.17 &
  0.48/0.76/0.93 &
  0.45/0.86/0.90 &
  8.03/0.00/0.12 &
  0.61/0.09 \\ \cline{2-10} 
              & whisper-tiny   & 0.54/0.74/0.85 & 0.45/0.87/0.93 & 7.94/0.19/0.21 & 0.69/0.16     & 0.43/0.76/0.91 & 0.45/0.83/0.90 & 8.20/-0.05/0.04 & 0.61/0.09     \\
              & whisper-base   & 0.57/0.73/0.85 & 0.45/0.86/0.93 & 7.94/0.17/0.17 & 0.69/0.17     & 0.43/0.78/0.91 & 0.45/0.83/0.90 & 8.24/-0.10/0.00 & 0.60/0.09     \\
              & whisper-small  & 0.57/0.72/0.89 & 0.45/0.88/0.93 & 7.72/0.26/0.20 & 0.69/0.17     & 0.46/0.76/0.93 & 0.46/0.85/0.90 & 8.10/-0.02/0.07 & 0.62/0.09     \\
              & whisper-medium & 0.61/0.73/0.89 & 0.45/0.88/0.93 & 7.70/0.27/0.23 & 0.69/0.20     & 0.47/0.79/0.93 & 0.45/0.85/0.90 & 8.08/-0.02/0.11 & 0.62/0.09     \\
              & GC-default     & 0.56/0.75/0.85 & 0.45/0.88/0.93 & 7.83/0.23/0.20 & 0.69/0.15     & 0.41/0.75/0.92 & 0.46/0.86/0.91 & 8.30/-0.08/0.02 & 0.62/0.09     \\
              & GC-med         & 0.56/0.74/0.88 & 0.45/0.87/0.92 & 7.71/0.28/0.23 & 0.69/0.15     & 0.47/0.78/0.91 & 0.45/0.84/0.91 & 8.04/0.06/0.13  & 0.62/0.09     \\
              & GC-video       & 0.56/0.75/0.85 & 0.45/0.88/0.93 & 7.83/0.23/0.20 & 0.69/0.15     & 0.41/0.75/0.92 & 0.46/0.86/0.91 & 8.31/-0.08/0.02 & 0.62/0.09     \\
              \hline
Manual &
  symptom truth &
  0.67/0.90/0.93 &
  0.52/0.88/0.93 &
  6.95/0.31/0.33 &
  0.69/0.13 &
  0.69/0.90/0.95 &
  0.55/0.90/0.95 &
  7.20/0.28/0.29 &
  0.64/0.09 \\ \hline
\multirow{7}{*}{GPT-4o} 
              & whisper-tiny   & 0.66/0.88/0.93 & \textbf{0.56}/0.86/\textbf{0.94} & 7.23/0.30/0.27 & 0.68/0.13     & 0.69/0.89/0.96 & 0.56/0.89/0.94 & 7.57/0.22/0.24  & \textbf{0.66}/0.09     \\
              & whisper-base   & 0.67/0.86/0.92 & 0.53/0.86/0.93 & 7.28/0.27/0.23 & 0.68/0.13     & 0.69/0.88/0.95 & 0.56/0.89/0.95 & 7.60/0.20/0.17  & 0.65/0.09     \\
              & whisper-small  & 0.67/0.87/0.93 & 0.53/\textbf{0.89}/0.94 & 6.98/0.34/0.32 & 0.68/0.12     & 0.66/0.88/0.95 & 0.55/0.90/0.95 & 7.26/0.29/0.27  & 0.64/0.09     \\
              & whisper-medium & \textbf{0.68}/0.88/0.92 & 0.52/0.88/0.93 & 6.99/0.32/0.33 & 0.68/0.12     & 0.69/0.89/0.95 & 0.54/0.90/0.95 & 7.25/0.28/0.27  & 0.64/0.09     \\
              & GC-default     & 0.66/0.88/0.93 & 0.56/0.89/0.93 & 7.13/0.31/0.24 & 0.69/0.12     & 0.66/0.88/0.96 & \textbf{0.57}/0.90/0.95 & 7.50/0.23/0.16  & 0.65/0.09     \\
              & GC-med         & 0.65/0.87/0.93 & 0.53/0.88/0.94 & 6.99/0.35/0.32 & 0.68/0.14     & 0.64/0.86/\textbf{0.97} & 0.54/0.90/0.94 & 7.23/0.32/0.28  & 0.65/0.09     \\
              & GC-video       & 0.65/0.88/0.93 & 0.56/0.89/0.93 & 7.13/0.31/0.24 & 0.69/0.12     & 0.66/0.88/0.96 & 0.57/0.90/0.95 & 7.50/0.23/0.16  & 0.65/0.09     \\
              \hline
\multirow{7}{*}{\textbf{EMSLlama(our)}} 
              & whisper-tiny   & 0.67/0.87/0.92 & 0.52/0.88/0.93 & 6.89/0.34/0.35 & \textcolor{blue}{\textbf{0.69}}/0.13     & 0.68/0.86/0.92 & 0.56/0.90/\textcolor{blue}{\textbf{0.95}} & 7.14/0.31/0.30  & 0.65/\textcolor{blue}{\textbf{0.09}}     \\
              & whisper-base   & 0.66/0.88/0.92 & 0.54/0.88/0.93 & 6.91/0.33/0.33 & 0.68/0.13     & 0.68/0.88/0.93 & 0.56/0.91/0.95 & 7.18/0.30/0.28  & 0.64/0.09     \\
              & whisper-small  & 0.66/\textcolor{blue}{\textbf{0.89}}/\textcolor{blue}{\textbf{0.93}} & 0.53/0.88/0.93 & 6.93/0.32/0.33 & 0.69/0.13     & 0.68/0.88/0.94 & 0.55/0.91/0.95 & 7.17/0.30/0.30  & 0.64/0.09     \\
              & whisper-medium & 0.66/0.89/0.93 & 0.54/0.88/0.93 & 6.94/0.31/0.31 & 0.68/0.13     & 0.68/\textcolor{blue}{\textbf{0.90}}/0.95 & 0.55/0.90/0.95 & 7.17/0.28/0.29  & 0.64/0.09     \\
              & GC-default     & 0.67/0.89/0.93 & 0.53/0.88/0.93 & \textcolor{blue}{\textbf{6.79}}/\textcolor{blue}{\textbf{0.38}}/\textcolor{blue}{\textbf{0.38}} & 0.68/0.13     & \textcolor{blue}{\textbf{0.69}}/0.89/0.93 & 0.55/\textcolor{blue}{\textbf{0.93}}/0.95 & \textcolor{blue}{\textbf{7.02}}/\textcolor{blue}{\textbf{0.36}}/0.38  & 0.64/0.09     \\
              & GC-med         & 0.65/0.85/0.92 & 0.54/0.88/0.93 & 6.96/0.31/0.31 & 0.68/\textcolor{blue}{\textbf{0.14}}     & 0.68/0.88/0.92 & 0.54/0.90/0.94 & 7.20/0.30/0.30  & 0.64/0.09     \\
              & GC-video       & 0.67/0.88/0.93 & 0.53/0.87/0.93 & 6.80/0.38/0.38 & 0.68/0.13     & 0.69/0.89/0.93 & 0.55/0.92/0.95 & 7.02/0.36/\textcolor{blue}{\textbf{0.39}}  & 0.64/0.09     \\
              \hline
\end{tabular}
}
\end{table*}

\subsubsection{End-to-end vital evaluation}
\label{sec:eval_e2e_vital_eval}


Table~\ref{tab:e2e_vital_eval} presents the end-to-end evaluation results when only analytics \#2 and \#3 are enabled, i.e., from video inputs to the four EMS inference tasks. As noted in Section~\ref{sec:design_analytics2_video_to_vitals}, TeleEMS does not aim to design new rPPG algorithms. Instead, we adopt the state-of-the-art TSCAN model for video-to-PPG conversion, and focus on evaluating how pre-arrival vitals derived from video signals can support EMS task inference. The preparation of post-arrival vitals (\textit{Vitals}$_0$) and pre-arrival vitals (\textit{Vitals}$_1$–\textit{Vitals}$_3$ and Video) is described in Section~\ref{sec:prearrival_data_prep}.

Our first finding is that while post-arrival vitals (\textit{Vitals}$_0$, containing six types of vitals) predict tasks with the highest accuracy, pre-arrival vitals still yield competitive performance. For instance, using only pre-arrival \textit{Vitals}$_1$, an LSTM model achieves 0.873 top-5 accuracy on the medicine type prescription task, close to the 0.904 accuracy of post-arrival vitals. This indicates that even with fewer vitals data available before EMTs arrive with professional equipment, pre-arrival vitals can already provide actionable decision support. Such results are critical in the pre-arrival phase, where EMTs en route must anticipate clinically effective interventions and guide bystanders in real time. This is extremely useful in emergency scenarios where streamed patient face videos are available but the audios are not available.

Our second finding is that the performance differences among the pre-arrival vitals (\textit{Vitals}$_1$–\textit{Vitals}$_3$ and Video) are very small. For example, in multi-task inference, the top-1 medicine type prescription accuracy varies only between 0.452 and 0.467 across three vitals models, i.e., RNN, LSTM, and GRU, with a maximum difference below 0.015. This consistency confirms that our proposed ``hooking'' process for generating pre-arrival video datasets (Section~\ref{sec:prearrival_data_prep}) produces valid and reliable evaluation data. While some degradation from \textit{Vitals}$_0$ to pre-arrival vitals is observed, especially for quantity prescription, the results demonstrate that TeleEMS can still inform pre-arrival decision-making through vitals-based analytics.

In practice, the evaluation of vitals-only data shows that even approximate vital signals derived from pre-arrival video can enable TeleEMS to bridge the gap between dispatch and EMT arrival, providing pre-arrival cues for preparation and improving the timeliness of care.

\begin{table*}[]
\caption{End-to-end EMS task accuracy using vitals-only data (video-only).}
\label{tab:e2e_vital_eval}
\scalebox{0.7}{
\begin{tabular}{lllcccccccc}
\hline
\multirow{3}{*}{\begin{tabular}[c]{@{}l@{}}Vitals\\ model\end{tabular}} &
  \multicolumn{2}{c}{\multirow{3}{*}{\begin{tabular}[c]{@{}c@{}}Vitals(HR)\\ data\end{tabular}}} &
  \multicolumn{4}{c}{Single task} &
  \multicolumn{4}{c}{Multiple task (4 simultaneous tasks)} \\
 &
  \multicolumn{2}{c}{} &
  Protocol &
  MedType &
  Quantity &
  Procedure(F1) &
  Protocol &
  MedType &
  Quantity &
  Procedure(F1) \\
 &
  \multicolumn{2}{c}{} &
  Top1/3/5 &
  Top1/3/5 &
  MSE/Pear/Spear &
  micro/macro &
  Top1/3/5 &
  Top1/3/5 &
  MSE/Pear/Spear &
  micro/macro \\ \hline
\multirow{5}{*}{RNN} &
  Post &
  Vitals0 &
  0.465/0.695/0.786 &
  0.520/0.822/0.909 &
  2.949/0.175/0.188 &
  0.619/0.111 &
  0.408/0.646/0.750 &
  0.496/0.805/0.903 &
  2.883/0.217/0.247 &
  0.607/0.092 \\ \cline{2-11} 
 &
  \multirow{4}{*}{Pre} &
  Vitals1 &
  0.342/0.632/0.722 &
  0.482/0.752/0.857 &
  3.034/0.060/0.081 &
  0.643/0.096 &
  0.317/0.590/0.688 &
  0.456/0.750/0.860 &
  3.005/0.071/0.050 &
  0.641/0.102 \\
 &
   &
  Vitals2 &
  0.333/0.738/0.810 &
  0.452/0.762/0.857 &
  7.951/-0.161/-0.161 &
  0.720/0.101 &
  0.357/0.738/0.833 &
  0.452/0.738/0.833 &
  7.719/-0.156/-0.223 &
  0.712/0.099 \\
 &
   &
  Vitals3 &
  0.357/0.643/0.810 &
  0.452/0.762/0.857 &
  7.881/-0.260/-0.281 &
  0.712/0.099 &
  0.381/0.643/0.810 &
  0.452/0.738/0.833 &
  7.410/-0.003/0.001 &
  0.712/0.099 \\
 &
   &
  Video &
  0.357/0.643/0.810 &
  0.452/0.762/0.857 &
  7.886/-0.239/-0.297 &
  0.712/0.099 &
  0.381/0.643/0.810 &
  0.452/0.738/0.833 &
  7.416/0.000/-0.069 &
  0.712/0.099 \\ \hline
\multirow{5}{*}{LSTM} &
  Post &
  Vitals0 &
  0.474/0.700/0.794 &
  0.512/0.818/0.904 &
  2.897/0.215/0.245 &
  0.646/0.096 &
  0.409/0.642/0.748 &
  0.492/0.800/0.898 &
  2.891/0.214/0.242 &
  0.611/0.092 \\ \cline{2-11} 
 &
  \multirow{4}{*}{Pre} &
  Vitals1 &
  0.350/0.636/0.725 &
  0.482/0.758/0.873 &
  3.020/0.089/0.068 &
  0.609/0.093 &
  0.288/0.578/0.674 &
  0.454/0.738/0.848 &
  3.019/0.080/0.052 &
  0.566/0.087 \\
 &
   &
  Vitals2 &
  0.357/0.714/0.833 &
  0.452/0.738/0.857 &
  7.989/-0.203/-0.215 &
  0.727/0.102 &
  0.333/0.762/0.833 &
  0.429/0.738/0.857 &
  7.998/-0.064/-0.106 &
  0.696/0.099 \\
 &
   &
  Vitals3 &
  0.381/0.643/0.810 &
  0.452/0.738/0.857 &
  7.575/-0.097/-0.066 &
  0.712/0.099 &
  0.333/0.762/0.810 &
  0.452/0.738/0.857 &
  7.780/0.050/-0.135 &
  0.712/0.099 \\
 &
   &
  Video &
  0.381/0.643/0.810 &
  0.452/0.762/0.857 &
  7.564/-0.086/-0.076 &
  0.712/0.099 &
  0.333/0.762/0.810 &
  0.452/0.738/0.857 &
  7.780/0.055/-0.138 &
  0.712/0.099 \\ \hline
\multirow{5}{*}{GRU} &
  Post &
  Vitals0 &
  0.485/0.714/0.804 &
  0.524/0.829/0.916 &
  2.889/0.220/0.244 &
  0.645/0.126 &
  0.426/0.659/0.760 &
  0.509/0.820/0.912 &
  2.848/0.238/0.276 &
  0.613/0.108 \\ \cline{2-11} 
 &
  \multirow{4}{*}{Pre} &
  Vitals1 &
  0.351/0.636/0.726 &
  0.480/0.759/0.870 &
  3.011/0.103/0.084 &
  0.609/0.093 &
  0.314/0.596/0.691 &
  0.467/0.754/0.869 &
  2.990/0.098/0.080 &
  0.639/0.096 \\
 &
   &
  Vitals2 &
  0.381/0.738/0.810 &
  0.452/0.762/0.833 &
  7.945/-0.180/-0.186 &
  0.727/0.102 &
  0.405/0.738/0.810 &
  0.452/0.762/0.833 &
  7.841/-0.164/-0.185 &
  0.720/0.101 \\
 &
   &
  Vitals3 &
  0.381/0.643/0.810 &
  0.452/0.762/0.833 &
  7.491/-0.030/-0.032 &
  0.712/0.099 &
  0.381/0.643/0.810 &
  0.452/0.762/0.857 &
  7.498/-0.089/-0.156 &
  0.712/0.099 \\
 &
   &
  Video &
  0.405/0.643/0.810 &
  0.452/0.762/0.833 &
  7.480/-0.003/-0.051 &
  0.712/0.099 &
  0.381/0.643/0.810 &
  0.452/0.762/0.857 &
  7.484/-0.062/-0.137 &
  0.712/0.099 \\ \hline
\end{tabular}
}
\end{table*}

\subsubsection{End-to-end text-vital evaluation}
\label{sec:eval_e2e_text_vital_eval}

Table~\ref{tab:e2e_text_vital_eval} reports the end-to-end evaluation when all three analytics are combined, i.e., both audio and video inputs flow into PreNet for joint inference of the four EMS tasks. This setup represents the full vision of TeleEMS: fusing text and vitals into a multimodal, multitask decision pipeline.

Two key observations emerge.
First, the accuracy values are consistently higher than those reported in Table~\ref{tab:e2e_vital_eval}, where only vitals data were used. This demonstrates that augmenting vitals with text enriches PreNet’s representation space and improves inference quality.
Second, when compared to the text-only pipeline in Section~\ref{sec:eval_e2e_text_eval} and Table~\ref{tab:e2e_text_eval}, the multimodal system can further outperform text-only inputs. Specifically, improvements appear in medicine type prescription (top-5 accuracy), medicine quantity prescription (MSE and Spearman), and procedure recommendation (F1 micro and F1 macro). These results indicate that text-vital fusion improves robustness for certain tasks, especially those involving quantities and actionable interventions. This further validates the significance of our data preparation effort, including the pre-arrival audio-video dataset construction and hooking process. While the dataset size for this end-to-end evaluation is limited (42 samples), the observed gains point to promising directions; future work on larger datasets will allow us to more systematically quantify the advantages of multimodal over unimodal pipelines.

Taken together, these findings highlight the real-world value of multimodal fusion for pre-arrival EMS. By jointly leveraging symptom text and vital data, TeleEMS can produce more accurate and reliable intervention recommendations. Such capabilities enable EMTs to anticipate patient-specific needs while en route, reducing critical delays in emergency care delivery and strengthening the operational impact of pre-arrival support.

\begin{table*}[]
\caption{End-to-end EMS task accuracy using text-vitals data (audio-video).}
\label{tab:e2e_text_vital_eval}
\scalebox{0.7}{
\begin{tabular}{llcccccccc}
\hline
Symptom                                  & Speech             & \multicolumn{4}{c}{Single task}                                    & \multicolumn{4}{c}{Multiple task (4 simultaneous tasks)}           \\
extraction                               & recognition        & Protocol       & MedType        & Quantity         & Procedure(F1) & Protocol       & MedType        & Quantity         & Procedure(F1) \\
normalization                            & model              & Top1/3/5       & Top1/3/5       & MSE/Pear/Spear   & micro/macro   & Top1/3/5       & Top1/3/5       & MSE/Pear/Spear   & micro/macro   \\ \hline
\multirow{8}{*}{None}                   & conversation truth & 0.55/0.81/0.95 & 0.52/0.83/0.90 & 7.77/0.01/0.16   & 0.72/0.11     & 0.55/0.79/0.93 & 0.52/0.83/0.83 & 7.54/0.03/0.08   & 0.66/0.10     \\ \cline{2-10} 
                                         & whisper-tiny       & 0.54/0.80/0.93 & 0.48/0.84/0.89 & 8.19/-0.10/-0.01 & \textbf{0.72}/0.11     & 0.53/0.74/0.89 & 0.48/0.81/0.83 & 7.70/-0.02/0.02  & 0.66/0.10     \\
                                         & whisper-base       & 0.54/0.82/0.92 & 0.50/0.86/0.90 & 8.16/-0.12/-0.01 & 0.72/0.11     & 0.53/0.75/0.89 & 0.47/0.81/0.83 & 7.71/-0.06/-0.01 & 0.65/0.10     \\
                                         & whisper-small      & 0.55/0.83/0.94 & 0.52/0.86/0.90 & 7.93/-0.02/0.06  & 0.72/0.11     & 0.54/0.77/0.91 & 0.50/0.82/0.83 & 7.55/0.03/0.06   & 0.65/0.10     \\
                                         & whisper-medium     & 0.54/0.83/0.94 & 0.51/0.85/0.90 & 7.89/-0.00/0.11  & 0.72/0.11     & 0.54/0.77/0.92 & 0.50/0.83/0.83 & 7.56/0.03/0.07   & 0.65/0.10     \\
                                         & GC-default         & 0.52/0.85/0.92 & 0.50/0.84/0.92 & 8.14/-0.08/-0.01 & 0.72/0.11     & 0.54/0.75/0.94 & 0.45/0.81/0.84 & 7.67/-0.03/0.00  & 0.65/0.10     \\
                                         & GC-med             & 0.50/0.82/0.92 & 0.51/0.85/0.91 & 7.97/-0.00/0.09  & 0.72/0.11     & 0.55/0.76/0.89 & 0.47/0.83/0.85 & 7.50/0.10/0.11   & 0.65/0.10     \\
                                         & GC-video           & 0.52/0.85/0.92 & 0.50/0.84/0.92 & 8.14/-0.08/-0.01 & 0.72/0.11     & 0.54/0.75/0.94 & 0.45/0.81/0.84 & 7.67/-0.03/0.00  & 0.65/0.10     \\
                                         \hline
Manual                                   & symptom truth      & 0.69/0.83/0.95 & 0.52/0.86/0.98 & 6.89/0.24/0.33   & 0.70/0.11     & 0.71/0.86/0.95 & 0.50/0.90/0.98 & 6.79/0.27/0.34   & 0.72/0.14     \\ \hline
\multirow{7}{*}{GPT4o}                  
                                         & whisper-tiny       & 0.67/0.85/0.95 & \textbf{0.56}/0.84/0.95 & 7.25/0.17/0.22   & 0.70/0.11     & 0.69/\textbf{0.87}/0.94 & 0.55/0.88/0.94 & 7.15/0.19/0.21   & 0.72/0.12     \\
                                         & whisper-base       & 0.65/0.82/0.94 & 0.55/0.84/0.95 & 7.28/0.15/0.18   & 0.70/0.11     & 0.68/0.86/0.95 & 0.53/0.89/0.95 & 7.15/0.18/0.19   & 0.71/0.12     \\
                                         & whisper-small      & 0.67/0.82/0.94 & 0.54/0.87/0.97 & 6.93/0.26/0.28   & 0.70/0.11     & 0.68/0.87/0.94 & 0.51/0.91/0.97 & 6.86/0.29/0.29   & 0.72/0.14     \\
                                         & whisper-medium     & 0.68/0.83/0.94 & 0.54/0.87/0.97 & 6.92/0.24/0.32   & 0.70/0.11     & 0.71/0.86/\textbf{0.96} & 0.50/0.91/0.98 & 6.87/0.27/0.31   & 0.72/0.14     \\
                                         & GC-default         & 0.65/0.82/\textbf{0.96} & 0.55/0.86/0.96 & 7.12/0.21/0.18   & 0.70/0.11     & 0.69/0.86/0.93 & 0.55/0.89/0.96 & 7.05/0.23/0.19   & 0.72/0.12     \\
                                         & GC-med             & 0.66/0.83/0.93 & 0.54/0.87/0.96 & 6.90/0.28/0.30   & 0.70/0.11     & 0.68/0.84/0.93 & 0.53/0.91/0.96 & 6.82/0.31/0.30   & 0.72/0.11     \\
                                         & GC-video           & 0.65/0.82/0.96 & 0.55/0.86/0.96 & 7.12/0.21/0.18   & 0.70/0.11     & 0.69/0.86/0.93 & \textbf{0.56}/0.89/0.96 & 7.05/0.23/0.19   & 0.72/0.13     \\
                                         \hline
\multirow{7}{*}{\textbf{EMSLlama(our)}} 
                                         & whisper-tiny       & 0.68/0.82/0.94 & 0.54/0.86/0.96 & 6.81/0.28/0.35   & 0.70/\textcolor{blue}{\textbf{0.11}}     & 0.70/0.85/0.92 & 0.49/\textcolor{blue}{\textbf{0.91}}/\textcolor{blue}{\textbf{0.98}} & 6.74/0.30/0.33   & \textcolor{blue}{\textbf{0.72}}/\textcolor{blue}{\textbf{0.14}}     \\
                                         & whisper-base       & 0.67/0.83/0.94 & 0.54/0.86/\textcolor{blue}{\textbf{0.98}} & 6.84/0.27/0.33   & 0.69/0.11     & 0.70/0.85/0.94 & 0.50/0.90/0.98 & 6.76/0.29/0.33   & 0.72/0.14     \\
                                         & whisper-small      & 0.68/0.83/0.94 & 0.53/0.86/0.97 & 6.87/0.25/0.33   & 0.70/0.11     & 0.70/0.86/0.94 & 0.50/0.89/0.98 & 6.76/0.29/0.33   & 0.72/0.14     \\
                                         & whisper-medium     & 0.68/0.83/0.94 & 0.55/0.86/0.98 & 6.89/0.24/0.32   & 0.69/0.11     & \textcolor{blue}{\textbf{0.71}}/0.86/0.95 & 0.50/0.90/0.98 & 6.78/0.27/0.33   & 0.72/0.14     \\
                                         & GC-default         & \textcolor{blue}{\textbf{0.69}}/\textcolor{blue}{\textbf{0.85}}/0.93 & 0.54/\textcolor{blue}{\textbf{0.88}}/0.98 & \textcolor{blue}{\textbf{6.76}}/\textcolor{blue}{\textbf{0.29}}/\textcolor{blue}{\textbf{0.37}}   & 0.69/0.11     & 0.71/0.86/0.95 & 0.49/0.88/0.97 & \textcolor{blue}{\textbf{6.62}}/\textcolor{blue}{\textbf{0.35}}/\textcolor{blue}{\textbf{0.40}}   & 0.72/0.14     \\
                                         & GC-med             & 0.66/0.82/0.92 & 0.54/0.85/0.97 & 6.89/0.24/0.31   & 0.69/0.11     & 0.69/0.85/0.94 & 0.52/0.90/0.97 & 6.79/0.28/0.31   & 0.72/0.14     \\
                                         & GC-video           & 0.68/0.85/0.93 & 0.54/0.87/0.98 & 6.77/0.29/0.37   & 0.69/0.11     & 0.71/0.86/0.95 & 0.49/0.88/0.97 & 6.63/0.35/0.40   & 0.72/0.14     \\
                                         \hline
\end{tabular}
}
\end{table*}

\section{Conclusions}
\label{sec:discuss_conclusion}

TeleEMS demonstrates how mobile live video analytics can fundamentally improve pre-arrival emergency medical services. By combining multiparty video streaming with real-time symptom extraction, vital estimation, and multimodal task inference, it transforms bystander video-audio inputs into actionable insights for dispatchers and EMTs. Our evaluation shows substantial gains over both GPT-4o and traditional baselines, with text–vital fusion further enhancing decision robustness. These results underscore TeleEMS as a step toward next-generation intelligent EMS systems, enabling faster, more reliable pre-arrival care.

\bibliographystyle{unsrt}
\bibliography{references}

\appendix
\section{Input prompts to GPT-4o}
\label{sec:input_prompt_gpt4o}

We use the following instructions, followed by 42 conversation transcriptions as the prompts to GPT-4o for the symptom extraction and normalization:

``Here are 40 conversation between an EMT and 911 callers. In each conversation, there may be more than one key medical terms to describe the patient's symptoms or disease. Can you extract all those key transcribed medical terms in each of these 40 conversations? You want to note that these are conversation audio transcriptions, so the medical terms may not be 100\% correct or normalized, but please extract and normalize them. If there is more than one medical terms in each conversation, please extract and normalize them all. Please concatenate the extracted symptoms into one sentence with an empty space. You don't need to show me the analysis/reasoning process, you don't need to provide the line number. Just give me the 40 lines of extracted and normalized symptom sentence.''

\end{document}